\documentclass[twoside]{article} 
\usepackage{amsmath,amssymb} 

\newtheorem{theorem}{Theorem}[section]
\newcommand{\demo}[1][$\!\!$]{\noindent\textbf{Proof }\textsl{#1}. }
\newcommand{\qed}{\hfill $\Box$\par\hfill\\}

\newtheorem{proposition}[theorem]{Proposition}
\newtheorem{lemma}[theorem]{Lemma}
\newtheorem{corollary}[theorem]{Corollary}
\newtheorem{definition}[theorem]{Definition}

\newtheorem{remark}{Remark}[section]
\newtheorem{example}{Example}[section]

\numberwithin{equation}{section}

\begin{document}

\title{Characterization of Infrared Catastrophe \\ 
by The Carleman Operator and Its Singularity}

\author{Masao Hirokawa\thanks{This work is supported by JSPS, 
Grant-in-Aid for Scientific Research (C) 18540180.} 
}

\date{\today}
\pagestyle{myheadings}
\markboth{M. Hirokawa}{IR Catastrophe and Carleman Operator}

\maketitle

\abstract{
This paper addresses some mathematical problems 
arising from the infrared (IR) catastrophe 
in quantum field theory. 
IR catastrophe is formulated and studied in operator theory, 
characterized by the Carleman operator. 
Non-existence of ground state under IR catastrophe 
is also investigated with the help of the characterization. 
The theory presented in this paper is applied 
to the Hamiltonian of the model describing a non-relativistic 
electron coupled with a quantum field of phonons or polaritons 
in the light of mathematics as well as solid state physics.   
}

\section{Introduction} 
\label{section:Intro}

The \textit{infrared (IR) catastrophe} comes up in a wide range 
of quantum field theory. 
Each sort of massless quanta makes a quantum field 
and has a possibility of its causing an individual 
IR divergence. 
In the concrete, the divergence of soft photons 
in quantum electrodynamics, 
the divergence of soft phonons in solid state physics, 
the divergence of soft gluons in quantum chromodynamics, 
etc. 
In this paper we formulate and handle IR 
catastrophe with a general framework of operator theory 
so that we can adapt our method to physical examples 
as much as possible. 
Another attempt from this point of view of 
general aspects was done in \cite{ahh2}. 
We consider Hamiltonians 
given by self-adjoint operators 
acting in a Hilbert space $\mathcal{F}$. 
Each Hamiltonian $H_{\mbox{\rm {\tiny QFT}}}$ 
represents the total energy of a physical system 
coupled with a quantum field. 
We suppose that $H_{\mbox{\rm {\tiny QFT}}}$ 
has \textit{IR singularity condition} \cite{ah2,ahh}. 
The order of the singularity depends on 
an individual model. 
So, some of models have IR catastrophe, some not. 
We express IR catastrophe by the divergence 
of the ground-state expectation 
$\left(\Psi_{\mbox{\rm {\tiny QFT}}}\, 
,\, N\Psi_{\mbox{\rm {\tiny QFT}}}\right)_{\mathcal{F}}$
of the total number of bosons, where $N$ 
is the boson \textit{number operator} acting in $\mathcal{F}$ 
and $\Psi_{\mbox{\rm {\tiny QFT}}}$ a \textit{ground state} 
of $H_{\mbox{\rm {\tiny QFT}}}$. 
The ground state $\Psi_{\mbox{\rm {\tiny QFT}}}$ 
is such an eigenvector of $H_{\mbox{\rm {\tiny QFT}}}$ 
that its eigenvalue is the lowest spectrum 
of $H_{\mbox{\rm {\tiny QFT}}}$. 
The so-called \textit{pull-through formula} 
\cite{fr2,jj,sch} is very useful for 
analyzing IR problems as well as for 
studying other problems in quantum field theory 
(see the literatures in the references of \cite{jj}). 
An idea to obtain the pull-through formula 
in operator theory was presented 
in \cite{hirokawa-nelson0} and it was completed 
in \cite{hirokawa-nelson}. 
Hiroshima showed in \cite[Theorem 2.9]{hiroshima} that 
we can derive the \textit{Carleman operator} \cite{weidmann} 
from the operator-theoretical pull-through formula, 
and then, he characterized a necessary and 
sufficient condition for the existence of ground state 
in the domain of $N^{1/2}$ by the Carleman operator in the case where 
IR catastrophe does not occur 
even if $H_{\mbox{\rm {\tiny QFT}}}$ has 
IR singularity condition. 
Conversely, we investigate IR catastrophe 
with the \textit{maximal} Carleman operator 
in this paper. 

Let us summarize our path and results here. 
In Section \ref{sec:setup}, we prepare some 
mathematical tools from quantum filed theory 
and we press ahead with our method through IR problems 
adopting Derezi\'{n}ski-G\'{e}rard's 
idea \cite{gerard-private}. 
Their idea is explained in Subsection \ref{subsec:dg} below. 
In Section \ref{sec:DPCOIRC}, we restate Hiroshima's 
\cite[Theorem 2.9]{hiroshima} 
and give another proof (Theorem \ref{theorem:CNB}). 
We characterize IR catastrophe by simple properties of 
the domain of the maximal Carleman operator 
(Theorems \ref{theorem:hougan}, 
\ref{lemma:zero}, and \ref{lemma:megane1}). 
In Section \ref{sec:IRCAGS}, we present 
a theorem on IR catastrophe 
(Theorem \ref{corollary:absence2-2}) and 
two theorems on absence of ground state 
(Theorems \ref{theorem:absence-base} 
and \ref{corollary:absence2-3}), 
using the simple domain properties and 
extending the notion of IR singularity condition 
(Definition \ref{def:IRSC}). 
Then, we can obtain Derezi\'{n}ski-G\'{e}rard's \cite[Lemma 2.6]{dg} 
and our \cite[Theorem 3.4]{ahh} as corollaries of 
one of the theorems 
(Corollaries \ref{proposition:absence1-1} and 
\ref{proposition:absence0}).  
We also prove that IR singularity condition 
prohibits $H_{\mbox{\rm {\tiny QFT}}}$ 
from making the mass gap. 
Namely, under IR singularity condition 
there is no spectral gap between the lowest 
spectrum (i.e., the \textit{ground state energy}) 
and the infimum of 
the essential spectrum of $H_{\mbox{\rm {\tiny QFT}}}$ 
(Theorem \ref{theorem:megane2}).  
Without giving a concrete form of $H_{\mbox{\rm {\tiny QFT}}}$, 
we assert all these results in a general framework 
so that our arguments are self-consistent 
in operator theory. 
The method presented in this paper enables us to analyze 
IR catastrophe by investigating the singularity of the 
maximal Carleman operator.

Let us briefly mention an application of our theory now. 
There have been many studies 
for the \textit{full} model 
in the so-called non-relativistic 
quantum electrodynamics under IR singularity condition 
(see \cite{bcfs,bfp,bfs2,gll,hiroshima1,hiroshima2,ll} 
and the literatures in their references). 
For this full model, 
there is no risk of its meeting IR catastrophe 
because it has local gauge invariance 
and thus it brings about the commutation relation, 
$i[H_{\mbox{\rm {\tiny QFT}}} , x] = v$, which 
cancels IR singularity, 
where $x$ and $v$ are the position and 
velocity of a non-relativistic electron respectively 
(see the explanation in \cite[p.212 and p.213]{hirokawa-e} 
about \cite{bfs2} and also Remark \ref{rem:PF}).  
Thus, Section \ref{sec:examples} of this paper addresses 
IR catastrophe for the Hamiltonian of the models describing 
a non-relativistic electron 
coupled with several types of phonons \cite{kittel} 
or polaritons \cite{IL}. 
This Hamiltonian is called the 
Pauli-Fierz (PF) Hamiltonian \cite{pf} 
by authors of \cite{dg,ggm,gerard}. 
They have made several pieces of painstaking research 
on its spectral theory, scattering theory, etc. 
Their PF Hamiltonian has the Fr\"{o}hlich 
interaction \cite{hfroehlich}, 
so it describes, for example, a non-relativistic 
electron in a polar crystal \cite{feynman,llp,lp} 
much better than the electron 
coupled with photons in quantum electrodynamics. 
From this point of view, existence of ground state 
and spectral properties of PF Hamiltonian was 
investigated in \cite{moeller}. 
Because of this physical situation, 
we call their PF Hamiltonian 
the \textit{Lee-Low-Pines (LLP) Hamiltonian} 
\cite[Eq.(1)]{lp} in Section \ref{sec:examples}. 
We apply our results to 
LLP Hamiltonian and investigate IR catastrophe for it 
in the light of mathematics 
as well as solid state physics. 
We presented in \cite{hirokawa-nelson} 
the possible physical mechanism for the situation 
that IR catastrophe occurs and then no ground 
state exists in $\mathcal{F}$. 
Namely, the size of the quasi particle dressed in 
the cloud of bosons is swelling 
as $\left(\Psi_{\mbox{\rm {\tiny QFT}}}\, 
,\, N\Psi_{\mbox{\rm {\tiny QFT}}}\right)_{\mathcal{F}}$ 
increases, and at last it becomes so large that 
we cannot observe the particle because the uncertainty 
of particle's position diverges in the ground state. 
That is when we lose any ground state 
in $\mathcal{F}$. 
In \cite{hirokawa-nelson} we showed 
this picture for the so-called 
Nelson Hamiltonian \cite{nelson} 
(i.e., the Gross Hamiltonian \cite{gross1,gross2}), 
directly adopting the idea of the spatial localization of 
ground state with exponential decay \cite{gll}.    
Based on this picture, 
as an application of our theory, 
we give a criterion for IR catastrophe 
for LLP Hamiltonian (Remark \ref{rem:UC} 
and Theorem \ref{theorem:criterion}). 
More precisely, let us set the dispersion relation 
$\omega(k)$ and the interaction function 
$1^{<\Lambda}(k)\rho(k)$ 
as $\omega(k) = |k|^{\mu}$ and $\rho(k) = |k|^{-\nu}$ 
for $\mu \ge 0$ and $\nu \in \mathbb {R}$, 
respectively,  
where $k \in  \mathbb{R}^{d}$ is the momentum of 
bosons, $\Lambda > 0$ a ultraviolet cutoff, and 
$1^{<\Lambda}(k)$ 
denotes the characteristic function of $|k| < \Lambda$. 
Obeying Spohn's result \cite{spohn}, if $\mu+\nu < d/2$, 
then LLP Hamiltonian has a ground state. 
On the other hand, IR catastrophe occurs for 
the ($3$-dimensional) Nelson Hamiltonian 
(i.e., $d=3$, $\mu=1$, and $\nu=1/2$) 
and then it does not have 
any ground state \cite{dg,hirokawa-nelson,lorinczi,panati}. 
Naturally, this result can be extended to 
LLP Hamiltonian with the condition, $\max\{ (\mu/2) + \nu, 
\mu + \nu - 1\} < d/2 < \mu + \nu$ 
(see Subsection \ref{subsec:dg} and 
Example \ref{example:dame}). 
Thus, we investigate the non-existence of 
ground state when $d, \mu, \nu$ are out of the regions. 
That is, we give a solution 
to the problem announced 
in \cite[Remark 2]{hirokawa-nelson}.  
Once we know that IR catastrophe occurs and thus 
there is no ground state in $\mathcal{F}$, 
we have to use non-Fock representation, 
which has been studied 
by \cite{arai,dg,fr2,hhs,panati,sasaki}.

\subsection{From Two 
Derezi\'{n}ski-G\'{e}rard's Ideas}
\label{subsec:dg}

When we estimate the ground-state expectation 
$\left(\Psi_{\mbox{\rm {\tiny QFT}}}\, 
,\, N\Psi_{\mbox{\rm {\tiny QFT}}}\right)_{\mathcal{F}}$ 
of the total number of bosons, 
it is convenient to use the pull-through formula 
in the equation: 
\begin{eqnarray}
\| N^{1/2}\Psi_{\mbox{\rm {\tiny QFT}}}\|_{\mathcal F}^{2} 
= \int_{{\mathbb R}^{d}}
\| a(k)\Psi_{\mbox{\rm {\tiny QFT}}}
\|_{\mathcal{F}}^{2}dk, 
\label{eq:establishment}
\end{eqnarray}
where $a(k)$ denotes the kernel of the so-called 
annihilation operator.
The method to establish 
Eq.(\ref{eq:establishment}) 
in operator theory is well known 
(see, e.g., 
\cite{hirokawa-nelson, hiroshima}, and also 
Proposition \ref{RIMS:cor-5.1}). 
On the other hand, when the integrand 
$\| a(k)\Psi_{\mbox{\rm {\tiny QFT}}}
\|_{\mathcal{F}}^{2}$ in Eq.(\ref{eq:establishment}) 
has a singularity at $k=0$, 
whether RHS of Eq.(\ref{eq:establishment}) converges 
is not certain. 
So, in such a case, we employ the following expression 
instead of Eq.(\ref{eq:establishment}): 
\begin{eqnarray}
\| N_{>\varepsilon}^{1/2}
\Psi_{\mbox{\rm {\tiny QFT}}}\|_{\mathcal F}^{2} 
= \int_{|k|>\varepsilon}\| a(k)\Psi_{\mbox{\rm {\tiny QFT}}}
\|_{\mathcal F}^{2}dk
\label{eq:establishment'}
\end{eqnarray}
for every $\varepsilon > 0$, where $N_{\varepsilon}$ 
is the number operator defined as the second quantization 
of $1^{>\varepsilon}$, the constant function $1(k) =1$ 
cut off within the radius of $\varepsilon$ from the origin. 
Thus, by taking $\varepsilon\to 0$ in 
Eq.(\ref{eq:establishment'}), 
we can investigate 
whether IR catastrophe occurs. 
This is the Derezi\'{n}ski-G\'{e}rard's 
idea \cite{gerard-private} 
which we adopt in our method, 
though they did not clearly write it 
in \cite{dg}.   
We establish Eq.(\ref{eq:establishment'}) 
in operator theory 
(Lemma \ref{lemma:establishment}). 
 
We note another Derezi\'{n}ski-G\'{e}rard's idea 
in \cite{dg} concerned with the decomposition 
of the plane wave. 
The typical model which represents 
the case where IR catastrophe occurs 
under IR singularity condition 
is the Nelson model. 
For the Nelson model the pull-through formula has 
the expression of 
$$a(k)\Psi_{\mbox{\rm {\tiny QFT}}} 
= 
- \left( H_{\mbox{\rm {\tiny QFT}}} 
- E_{0}(H_{\mbox{\rm {\tiny QFT}}}) + \omega(k)\right)^{-1}
\left(1^{<\Lambda}(k)\rho(k)
e^{-ikx}\right)\Psi_{\mbox{\rm {\tiny QFT}}},$$
where $E_{0}(H_{\mbox{\rm {\tiny QFT}}})$ 
is the ground state energy of 
$H_{\mbox{\rm {\tiny QFT}}}$. 
We note that this formula should be mathematically 
established in a certain sense as in 
\cite{ahh2,bd,gerard,hirokawa-nelson}. 
Because the domain of $a(k)$ is so narrow that 
$a(k)$ is not closable (see e.g., 
\cite[Remark1]{hirokawa-nelson}) 
when regarded as an operator, 
and moreover, 
$a(k)\Psi_{\mbox{\rm {\tiny QFT}}}$ may have 
the singularity at $k=0$ now. 
Another Derezi\'{n}ski-G\'{e}rard's idea 
in \cite[Lemma 2.2]{dg} is 
the simple decomposition $e^{-ikx} 
= 1 + (e^{-ikx} -1)$. 
Following their idea, $a(k)\Psi_{\mbox{\rm {\tiny QFT}}}$ 
can be decomposed into the dipole-approximated term 
$J_{\mathrm{dip}}(k)\Psi_{\mbox{\rm {\tiny QFT}}}$ 
and the error term 
$J_{\mathrm{err}}(k)\Psi_{\mbox{\rm {\tiny QFT}}}$, 
i.e., 
$a(k)\Psi_{\mbox{\rm {\tiny QFT}}} 
= J_{\mathrm{dip}}(k)\Psi_{\mbox{\rm {\tiny QFT}}} 
+ J_{\mathrm{err}}(k)\Psi_{\mbox{\rm {\tiny QFT}}}$. 
We know 
$J_{\mathrm{err}}(k)\Psi_{\mbox{\rm {\tiny QFT}}}$ is 
IR-safe 
(i.e., $J_{\mathrm{err}}(\cdot)\Psi_{\mbox{\rm {\tiny QFT}}} 
\in  L^{2}(\mathbb{R}^{d} ; \mathcal F)$) 
for the Nelson model ($d=3$, $\mu=1$, and $\nu=1/2$) 
by using $|e^{-ikx} - 1| \le |k||x|$.  
Here, of course, showing this square integrability 
usually requires that 
$\Psi_{\mbox{\rm {\tiny QFT}}} \in  D(|x|)$ 
whenever $\Psi_{\mbox{\rm {\tiny QFT}}}$ exists. 
Obeying this method, 
to show the error term 
$J_{\mathrm{err}}(k)
\Psi_{\mbox{\rm {\tiny QFT}}}$ is IR-safe 
for LLP Hamiltonian, 
the dimension $d$ is usually 
restricted from below as 
$\mu + \nu -1 < d/2$. 
In general, it is difficult to show 
that $J_{\mathrm{err}}(k)
\Psi_{\mbox{\rm {\tiny QFT}}}$ is IR-safe 
for LLP Hamiltonian 
without this restriction. 
Under the restriction, whether IR catastrophe occurs 
(i.e., whether RHS of Eq.(\ref{eq:establishment}) 
diverges) depends on 
whether the dipole-approximated term 
$J_{\mathrm{dip}}(k)\Psi_{\mbox{\rm {\tiny QFT}}}$ 
is IR-divergent (i.e., $J_{\mathrm{dip}}(\cdot)
\Psi_{\mbox{\rm {\tiny QFT}}} 
\notin  L^{2}(\mathbb{R}^{d} ; \mathcal F)$). 
Indeed the error term 
$J_{\mathrm{err}}(k)\Psi_{\mbox{\rm {\tiny QFT}}}$ 
becomes IR-safe under the restriction, 
but the following question arises. 
How can we prove IR catastrophe and 
non-existence of ground state 
when we do not know whether the error term 
$J_{\mathrm{err}}(k)\Psi_{\mbox{\rm {\tiny QFT}}}$ 
is IR-safe? 
Namely, how can we remove the restriction on $d$ 
from below? 
This question was stated in \cite[Remark 2]{hirokawa-nelson}. 
This paper addresses this question. 

\section{Set-ups in Mathematics}
\label{sec:setup}

In this section we prepare some tools from 
mathematics for quantum field theory 
and give our Hamiltonian $H_{\mbox{\rm {\tiny QFT}}}$. 
Once we obtain the maximal Carleman operator and 
its domain properties in Section \ref{sec:DPCOIRC}, 
the almost only thing we do is to analyze 
the Carleman operator and its singularity. 

\subsection{Preliminaries}

Let $X = (X, \mathcal{A}, \mu)$ be a $\sigma$-finite 
measurable space. 
Let us denote by $X^{n}$ 
$n$-fold Cartesian product of $X$. 
The measure for $X^{n}$ is naturally given 
by $d\mu^{n}(k_{1},\cdots,k_{n}) := 
d\mu(k_{1})\otimes\cdots\otimes d\mu(k_{n})$. 
Thus, we define the boson Fock space ${\mathcal F}_{\mathrm b}(L^{2}(X))$ 
over $L^{2}(X) := L^{2}(X, \mathcal{A}, \mu)$ by 
\begin{eqnarray*}
{\mathcal F}_{\mathrm b}(L^{2}(X)) := 
\bigoplus_{n=0}^{\infty}\otimes_{\mathrm s}^{n}
L^{2}(X).  
\label{eq:FockSpace}
\end{eqnarray*}
Here, $\otimes_{\mathrm s}^{n}L^{2}(X)$ 
is the $n$-fold symmetric tensor product of 
$L^{2}(X)$ for $n \in  {\mathbb N}$ 
with convention $\otimes_{\mathrm s}^{0}L^{2}(X) 
:= {\mathbb C}$. 
For $\psi \in  {\mathcal F}_{\mathrm b}(L^{2}(X))$, 
we use the following notation: 
\begin{eqnarray*}
\psi = \oplus\sum_{n=0}^{\infty}\psi^{(n)}, 
\quad 
\psi^{(n)} \in  \otimes_{\mathrm s}^{n}L^{2}(X)\,\,\,;\,\,\, 
n \in  \left\{ 0\right\}\cup{\mathbb N}. 
\label{eq:notation0}
\end{eqnarray*}
We often abbreviate ${\mathcal F}_{\mathrm b}(L^{2}(X))$ 
to ${\mathcal F}_{X}$ for simplicity in this paper, i.e., 
$$
{\mathcal F}_{X}^{\,} := 
{\mathcal F}_{\mathrm b}(L^{2}(X)). 
$$ 
We employ the standard norm $\|\quad
\|_{\mathcal{F}_{X}}$ in $\mathcal{F}_{X}$. 
We denote by $\|\quad\|_{\mathcal{V}}$ 
the norm of a Hilbert space $\mathcal{V}$, 
induced its inner product, 
throughout this paper. 

For each $n \in \{ 0\}\cup {\mathbb N}$ 
and every $f \in  L^{2}(X)$, 
we define an operator 
$a_{X}^{\,}(f) : \otimes_{\mathrm s}^{n}L^{2}(X) \ni 
\psi^{(n+1)} \mapsto (a_{X}^{\,}(f)\psi)^{(n)} 
\in \otimes_{\mathrm s}^{n}L^{2}(X)$ by 
\begin{eqnarray*}
\left( a_{X}^{\,}(f)\psi\right)^{(n)}(k_{1}, \cdots, k_{n}) 
:= \sqrt{n+1}\int_{X}\overline{f(k)}\psi^{(n+1)}
(k, k_{1}, \cdots, k_{n})d\mu(k). 
\label{eq:smeared-annihilation-op-0}
\end{eqnarray*}
We can extend $a_{X}^{\,}(f)$ to a closed 
operator acting in 
${\mathcal F}_{X}$ as 
\begin{eqnarray*}
&{}& 
a_{X}^{\,}(f)\psi := 
\oplus\sum_{n=0}^{\infty}
\left( a_{X}^{\,}(f)\psi\right)^{(n)}, \\ 
&{}& 
D(a_{X}^{\,}(f)) := 
\left\{
\psi \in {\mathcal F}_{X}^{}
\Bigg|\, 
\sum_{n=0}^{\infty}
\|
\left( a_{X}^{\,}(f)\psi\right)^{(n)}
\|_{\otimes_{\mathrm s}^{n}L^{2}(X)}^{2} 
< \infty
\right\}. 
\label{eq:smeared-annihilation-op}
\end{eqnarray*}
We call $a_{X}^{\,}(f)$ the {\it annihilation operator}. 
Since we can regard it as an operator-valued distribution, 
symbolically we often write it as 
\begin{eqnarray*}
a_{X}^{\,}(f) 
= \int_{X}\overline{f(k)}a_{X}^{\,}(k)d\mu(k)
\label{eq:kernel-annihilation}
\end{eqnarray*}
with a kernel $a_{X}^{\,}(k)$ of the annihilation operator. 
We define the {\it creation operator} $a_{X}^{\dagger}(f)$ 
for every $f \in  L^{2}(X)$ by $a_{X}^{\,}(f)^{*}$, 
the adjoint operator of $a_{X}^{\,}(f)$, i.e.,  
$a_{X}^{\dagger}(f) := a_{X}^{\,}(f)^{*}$. 
The kernel of $a_{X}^{\dagger}(f)$ is denoted 
as $a^{\dagger}_{X}(k)$ frequently.

Let $T$ be every closable operator 
densely defined in $L^{2}(X)$. 
For $n \in  \{ 0\}\cup 
\mathbb{N}$ we set $T^{(0)}$ as 
$T^{(0)} := 0$ and define 
$T^{(n)} : \otimes^{n}_{\mathrm {s}}
L^{2}(X) \to \otimes^{n}_{\mathrm {s}}
L^{2}(X)$ by  
\begin{eqnarray*}
T^{(n)} := 
\overline{\sum_{j=1}^{n}
I\otimes\cdots\otimes I\otimes 
\mathop{\mathop{T}_{\smallfrown}
}_{\mbox{\rm {\tiny$j$-th}}}
\otimes I 
\otimes\cdots\otimes I}.  
\label{eq:T(n)}
\end{eqnarray*}
We denote by $\overline{S}$ the closure of 
a closable operator $S$.  
Then, we define an operator 
$d\Gamma_{X}(T)$ acting in 
$\mathcal{F}_{X}$ by 
\begin{eqnarray*}
d\Gamma_{X}(T) := 
\bigoplus_{n=0}^{\infty}
T^{(n)}. 
\label{eq:2nd-quantization}
\end{eqnarray*}
We call $d\Gamma_{X}(T)$ 
{\it second quantization of $T$}. 
For the second quantization 
the following facts are well known: 

\begin{proposition}
\label{proposition:2nd-quantization} 
\qquad 
\begin{description}
\item[(i)] If $T \ne 0$, then $d\Gamma_{X}(T)$ 
is unbounded.
\item[(ii)] If $T$ is self-adjoint, then 
$d\Gamma_{X}(T)$ is also self-adjoint. 
\item[(iii)] Let $T$ be non-negative, injective, 
and self-adjoint. Then, for every 
$f \in  D(T^{-1/2})$ 
\begin{eqnarray*}
D(d\Gamma_{X}(T)^{1/2}) 
\subset 
D(a_{X}(f))\cap D(a^{\dagger}_{X}(f)). 
\label{eq:A-2-24-1}
\end{eqnarray*}
In addition, for every $\psi \in  
D(d\Gamma_{X}(T)^{1/2})$
\begin{eqnarray*}
\| a_{X}(f)\psi\|_{\mathcal{F}_{X}} 
&\le&  
\| T^{-1/2}f\|_{L^{2}(X)}
\|d\Gamma_{X}(T)^{1/2}\psi\|_{\mathcal{F}_{X}}, 
\label{eq:A-2-24-2} \\ 
\nonumber 
\| a^{\dagger}_{X}(f)\psi\|_{\mathcal{F}_{X}} 
&\le&   
\| T^{-1/2}f\|_{L^{2}(X)}
\|d\Gamma_{X}(T)^{1/2}\psi\|_{\mathcal{F}_{X}} \\ 
&{}&   
+ 
\| f\|_{L^{2}(X)}
\|\psi\|_{\mathcal{F}_{X}}. 
\label{eq:A-2-24-3} 
\end{eqnarray*}
\end{description}

\end{proposition}

\quad

Let $1$ stand for the multiplication operator 
of the constant function $1(k) \equiv 1$ 
of $k \in  X$ now. 
Then, we define an operator $N_{X}^{\,}$ 
acting in $\mathcal{F}_{X}^{\,}$ by 
$$N_{X}^{\,} := 
d\Gamma_{X}^{\,}(1).$$

Let $X$ be able to be 
decomposed into the disjoint union of 
$X_{1}$ and $X_{2}$, i.e., $X = X_{1}\cup X_{2}$ 
and $X_{1}\cap X_{2} = \emptyset$. 
Then, $L^{2}(X)$ is also decomposed into 
the direct sum of $L^{2}(X_{1})$ and 
$L^{2}(X_{2})$, i.e., 
$L^{2}(X) = L^{2}(X_{1})\bigoplus L^{2}(X_{2})$. 
The following proposition 
is known: 

\begin{proposition}
\label{A-theo-4-55-56}
There is a unique unitary operator $U : 
\mathcal{F}_{X}^{\,} \equiv 
\mathcal{F}_{\mathrm b}(L^{2}(X)) \to 
\mathcal{F}_{X_{1}}^{\,}\otimes 
\mathcal{F}_{X_{2}}^{\,} \equiv 
\mathcal{F}_{\mathrm b}(L^{2}(X_{1}))\otimes 
\mathcal{F}_{\mathrm b}(L^{2}(X_{2}))$ 
such that 
\begin{description}
\item[(i)] For the individual Fock vacuum, 
$\Omega_{X}^{\,} \in {\mathcal F}_{X}^{\,}, 
\, \Omega_{X_{1}}^{\,} \in {\mathcal F}_{X_{1}}^{\,}$, 
and $\Omega_{X_{2}}^{\,} \in 
{\mathcal F}_{X_{2}}^{\,}$,  
$$U\Omega_{X}^{\,} = 
\Omega_{X_{1}}^{\,}\otimes\Omega_{X_{2}}^{\,}.$$ 
\item[(ii)] For the decomposition $h = h_{1}\oplus h_{2}$\, 
($h \in  L^{2}(X)$, $h_{j} \in  L^{2}(X_{j})$, $j =1,2$), 
$$Ud\Gamma_{X}^{\,}(h) = 
\overline{d\Gamma_{X_{1}}^{\,}(h_{1})\otimes I + 
\otimes d\Gamma_{X_{2}}^{\,}(h_{2})}. 
$$
\end{description}
\end{proposition}

Let ${\mathcal V}$ be a separable Hilbert space. 
Then, for each $n \in  {\mathbb N}$ 
we define the Hilbert space 
$L_{\mathrm {sym}}^{2}(X^{n} ; {\mathcal V})$ 
of all square-integrable, ${\mathcal V}$-valued, symmetric 
functions: 
\begin{eqnarray*}
L_{\mathrm {sym}}^{2}(X^{n} ; {\mathcal V}) 
&:=& 
\Biggl\{
f : X^{n} \to {\mathcal V}\,\mbox{is measurable} 
\Bigg|\, \mbox{for each $\sigma \in  \mathfrak{S}_{n}$} \\ 
&{}& \qquad\qquad 
f(k_{1}, \cdots,k_{n}) = 
f(k_{\sigma(1)}, \cdots,k_{\sigma(n)}) \\ 
&{}& \qquad\mbox{and}\,  
\int_{X^{n}}\| f(k_{1},\cdots,k_{n})\|_{\mathcal V}^{2}
d\mu^{n}(k_{1},\cdots,k_{n}) < \infty 
\Biggr\},  
\end{eqnarray*}
where $\mathfrak{S}_{n}$ denotes the permutation group 
of all permutations of $\{ 1, \cdots, n\}$, i.e., 
$\mathfrak{S}_{n} \ni \sigma$ is a bijective map 
from $\{ 1, \cdots, n\}$ to itself. 
We say $f : X^{n} \to {\mathcal V}$ is measurable 
if $( v , f(\cdot))_{\mathcal V} : X^{n} \to  {\mathbb C}$ 
is measurable for every $v \in {\mathcal V}$.

The following proposition is well known: 

\begin{proposition}
\label{proposition:070630-1}
The two spaces, 
$\mathcal{V}\otimes\mathcal{F}_{X}$ and 
$\bigoplus_{n=0}^{\infty}
L_{\mathrm{sym}}^{2}(X^{n};\mathcal{V})$, 
are unitarily equivalent. 
Namely, there is a unitary operator 
$U_{\mathcal V} : {\mathcal V}\otimes 
{\mathcal F}_{X} \to 
\bigoplus_{n=0}^{\infty}
L_{\mathrm {sym}}^{2}(X^{n} ; {\mathcal V})$ 
with convention $L_{\mathrm {sym}}^{2}(X^{0} ; {\mathcal V}) 
:= {\mathcal V}$. 
\end{proposition}

Through this unitary transformation $U_{\mathcal{V}}$, 
for every $\Psi \in {\mathcal F}$ we denote  
$U_{\mathcal V}\Psi$ by 
$\Psi_{\mathcal V}^{\,}$, i.e., 
$\Psi_{\mathcal V}^{\,} := U_{\mathcal V}\Psi$. 
Moreover, $\Psi_{\mathcal V}^{\,}$ 
is often expressed as   
\begin{eqnarray}
\Psi_{\mathcal V}^{\,}  
&=& \oplus\sum_{n=0}^{\infty}\Psi_{\mathcal V}^{(n)} 
= \Psi_{\mathcal V}^{(0)}\oplus
\Psi_{\mathcal V}^{(1)}\oplus\cdots\oplus
\Psi_{\mathcal V}^{(n)}\oplus\cdots, 
\label{eq:V-rep} \\ 
\nonumber 
&{}& \qquad\qquad\qquad
\Psi_{\mathcal V}^{(n)} \in  
L^{2}_{\mathrm {sym}}(X^{n} ; {\mathcal V}),\,\,\, 
n \in  \{0\}\cup{\mathbb N}.
\end{eqnarray}
Therefore, the norm $\|\Psi\|_{{\mathcal F}_{X}^{\,}}$ 
has the following expression: 
\begin{eqnarray*}
\nonumber 
\|\Psi\|_{\mathcal{F}_{X}}^{2} 
&=& 
\|\Psi_{\mathcal V}^{(0)}\|_{\mathcal V}^{2} 
+ \sum_{n=1}^{\infty}
\|\Psi_{\mathcal V}^{(n)}
\|_{L^{2}(X^{n};{\mathcal V})}^{2} \\ 
&=& \|\Psi_{\mathcal V}^{(0)}\|_{\mathcal V}^{2} 
+ \sum_{n=1}^{\infty}
\int_{X^{n}}
\|\Psi_{\mathcal V}^{(n)}(k_{1}, \cdots, k_{n})\|_{\mathcal V}^{2}
d\mu^{n}(k_{1},\cdots,k_{n}). 
\end{eqnarray*}

Here we give the generalization of 
\cite[Corollary 5.1]{hirokawa-nelson} 
together with its proof: 

\begin{proposition}
\label{RIMS:cor-5.1}
Let $\{ f_{\ell}^{\,}\}_{\ell=1}^{\infty}$ be 
an arbitrary complete orthonormal system 
of $L^{2}(X)$. Then, 
$$
\| I\otimes N_{X}^{1/2}\Psi
\|_{{\mathcal V}\otimes
{\mathcal F}_{X}^{\,}}^{2} 
= \sum_{\ell=1}^{\infty}
\| I\otimes a_{X}^{\,}(f_{\ell}^{\,})\Psi
\|_{{\mathcal V}\otimes
{\mathcal F}_{X}^{\,}}^{2} 
$$
for every $\Psi \in 
D(I\otimes N_{X}^{1/2})$. 
\end{proposition}

\demo Let $\Psi \in 
D(I\otimes N_{X}^{1/2})$.
Then, by the definition of 
the annihilation operator 
and Proposition \ref{proposition:070630-1}, 
for each $M \in  {\mathbb N}$ we have 
\begin{eqnarray}
\nonumber 
&{}&
\sum_{\ell=1}^{M}
\| I\otimes a_{X}^{\,}
(f_{\ell}^{\,})
\Psi\|_{{\mathcal V}\otimes 
{\mathcal F}_{X}^{\,}}^{2} 
= 
\sum_{\ell=1}^{M}
\| a_{\mathcal V}(f_{\ell}^{\,})
\Psi_{\mathcal V}\|_{\bigoplus_{n}
L_{\mathrm {sym}}^{2}
(X^{n};{\mathcal V})}^{2} \\  
&=& 
\sum_{n=0}^{\infty}
(n+1)\int_{X^{n}}
\Psi_{M,\varepsilon}^{(n)}
(k_{1}, \cdots, k_{n})
d\mu^{n}(k_{1},\cdots, k_{n}), 
\label{eq:new-1-0}
\end{eqnarray}
where 
\begin{eqnarray*}
\Psi_{M,\varepsilon}^{(n)}
(k_{1}, \cdots, k_{n}) 
:= 
\sum_{\ell=1}^{M} 
\Bigg\| 
\int_{X} 
\overline{f_{\ell}^{\,}(k)}\, 
\Psi_{\mathcal V}^{(n+1)}
(k, k_{1}, \cdots, k_{n})d\mu(k)
\Bigg\|_{\mathcal V}^{2}. 
\label{eq:new-2-0}
\end{eqnarray*}
Let $\{ e_{p}\}_{p=1}^{\infty}$ be an arbitrary 
complete orthonormal system of ${\mathcal V}$. 
Then, we have 
\begin{eqnarray*}
&{}& 
\Psi_{M,\varepsilon}^{(n)}
(k_{1}, \cdots, k_{n}) \\ 
&=&  
\sum_{\ell=1}^{M}\sum_{p=1}^{\infty}
\Bigg| \left( e_{p}\, ,\, 
\int_{X} 
\overline{f_{\ell}^{\,}(k)} 
\Psi_{\mathcal V}^{(n+1)}
(k, k_{1}, \cdots, k_{n})d\mu(k) 
\right)_{\mathcal V}
\Bigg|^{2} \\ 
&=&  
\sum_{\ell=1}^{M}\sum_{p=1}^{\infty}
\Bigg| 
\int_{X}
\left(f_{\ell}^{\,}(k) e_{p}\, ,\, 
\Psi_{\mathcal V}^{(n+1)}
(k, k_{1}, \cdots, k_{n})
\right)_{\mathcal V}d\mu(k)
\Bigg|^{2}. 
\end{eqnarray*}
We note here that 
$\Psi_{\mathcal V}^{(n+1)}
(\cdot, k_{1}, \cdots, k_{n}) 
\in L^{2}(X; {\mathcal V})$ 
for a.e. $(k_{1}, \cdots, k_{n})$. 
Moreover, $\{ f_{\ell}^{\,}e_{p}
\}_{\ell,p=1}^{\infty}$ 
makes a complete orthonormal system of 
$L^{2}(X; {\mathcal V})$. 
Hence it follows from Bessel's inequality that 
$$
\Psi_{M,\varepsilon}^{(n)}
(k_{1}, \cdots, k_{n}) 
\le 
\|\Psi^{(n+1)}_{\mathcal V}
(\cdot, k_{1}, \cdots, k_{n})\|_{
L^{2}(X; {\mathcal V})
}^{2}.  
$$ 
Thus, we have 
\begin{eqnarray}
\Psi_{M,\varepsilon}^{(n)}
(k_{1}, \cdots, k_{n}) 
\nearrow 
\int_{X} 
\|\Psi^{(n+1)}_{\mathcal V}(k,k_{1},\cdots,k_{n})
\|_{\mathcal V}^{2}d\mu(k)
\label{eq:new-3-0}
\end{eqnarray}
as $M\to\infty$. 
Applying Lebesgue's monotone convergence theorem 
and Fubini's theorem to Eqs.(\ref{eq:new-1-0}) 
and (\ref{eq:new-3-0}), 
we reach the conclusion: 
\begin{eqnarray*}
&{}& 
\sum_{\ell=1}^{\infty}
\| I\otimes a_{X}^{\,}(f_{\ell}^{\,})
\Psi\|_{{\mathcal V}\otimes
{\mathcal F}_{X}^{\,}}^{2} \\ 
&=& 
\sum_{n=0}^{\infty}(n+1)
\int_{X^{n+1}} 
\|\Psi^{(n+1)}(k_{1},\cdots,k_{n+1})
\|_{\mathcal V}^{2}
d\mu^{n}(k_{1},\cdots, k_{n+1}) \\ 
&=& 
\| I\otimes N_{X}^{1/2}
\Psi\|_{{\mathcal V}\otimes
{\mathcal F}_{X}^{\,}}^{2} .
\end{eqnarray*}
\qed

As a special case of Proposition 
\ref{RIMS:cor-5.1}, 
namely we only have to take the case 
where ${\mathcal V} = {\mathbb C}$, 
we have

\begin{corollary}
\cite[Proposition 5.1]{hirokawa-nelson}
\label{RIMS:prop-5.1}
Let $\{ f_{\ell}^{\,}\}_{\ell=1}^{\infty}$ be 
an arbitrary complete orthonormal system of $L^{2}(X)$. 
Then, 
$$
\| N_{X}^{1/2}\psi
\|_{{\mathcal F}_{X}^{\,}}^{2} 
= \sum_{\ell=1}^{\infty}
\| a_{X}^{\,}(f_{\ell}^{\,})\psi
\|_{{\mathcal F}_{X}^{\,}}^{2} 
$$
for every $\psi \in  D(N_{X}^{1/2})$. 
\end{corollary}

\subsection{The Total Hamiltonian $H_{\mbox{\rm {\tiny QFT}}}$}

Let us give the state space of the physical system 
represented by a separable, 
complex Hilbert space ${\mathcal H}$.   
Only when $X = {\mathbb R}^{d}$, 
we use the following abbreviation:  
$${\mathcal F}_{\mathrm b} 
:= {\mathcal F}_{{\mathbb R}^{d}}^{\,} 
\equiv {\mathcal F}_{\mathrm b}
(L^{2}({\mathbb R}^{d})). 
$$ 
Corresponding to this abbreviation, we abbreviate 
$a_{{\mathbb R}^{d}}(f)$, 
$a_{{\mathbb R}^{d}}^{\dagger}(f)$, 
and $d\Gamma_{{\mathbb R}^{d}}(h)$ 
to $a_{\mathrm b}(f)$, 
$a_{\mathrm b}^{\dagger}(f)$, 
and $d\Gamma_{\mathrm b}(h)$, 
respectively:
$$
a_{\mathrm b}(f) := a_{{\mathbb R}^{d}}(f), 
\quad 
a_{\mathrm b}^{\dagger}(f) 
:= a_{{\mathbb R}^{d}}^{\dagger}(f), 
\quad 
d\Gamma_{\mathrm b}(h) := d\Gamma_{{\mathbb R}^{d}}(h). 
$$ 
In particular we often use the notation 
$N_{\mathrm b}$ for $d\Gamma_{\mathrm b}(1)$, 
i.e., 
$$N_{\mathrm b} := 
d\Gamma_{\mathrm b}(1).$$

The total state space of the physical system 
coupled with the Bose field is given by 
the tensor product of the two Hilbert spaces:  
$$
{\mathcal F} 
:= {\mathcal H}\otimes {\mathcal F}_{\mathrm b}. 
$$

Let $A$ be a self-adjoint operator 
acting in ${\mathcal H}$ 
bounded from below. 
We suppose the following idealization 
for the dispersion relation $\omega(k)$ 
because we are interested in IR behavior 
around $k=0$. 
Let $\omega : {\mathbb R}^{d} \longrightarrow 
\left.\left[ 0\, ,\, 
\infty\right.\right)$ be a continuous function 
such that $0 < \omega(k) < \infty$ for every 
$k \in  \mathbb{R}^{d}\setminus\{ 0\}$ and 
$\inf_{|k|>\varepsilon}\omega(k) > 0$ for every 
$\varepsilon > 0$. 
The unperturbed Hamiltonian of our model 
is defined by 
\begin{eqnarray}
H_{0} := A\otimes I 
+ I\otimes d\Gamma_{\mathrm b}(\omega) 
\label{eq:Free}
\end{eqnarray}
with domain $D(H_{0}) := D(A\otimes I)\cap 
D(I\otimes d\Gamma_{\mathrm b}(\omega)) \subset 
{\mathcal F}$, where $I$ denotes identity operator 
and $D(S)$ the domain of an operator $S$. 
The operator $H_{0}$ is self-adjoint and bounded from below.

We suppose that our total Hamiltonian 
has the form: 
$$H_{{\mbox{\rm {\tiny QFT}}}} 
= H_{0} + H_{\mathrm I},
$$  
and \textit{we always assume $H_{{\mbox{\rm {\tiny QFT}}}}$ 
to be a self-adjoint operator 
acting in ${\mathcal F}$} in this paper and 
we suppose it to describe our model of the physical system 
coupled with the quantum field. 
Here $H_{\mathrm I}$ is the interaction 
Hamiltonian. 

Let $\mathrm{ker}(S)$ stand for the kernel 
of an operator $S$, i.e., 
$$
\mathrm{ker}(S) 
:= \left\{
\Psi \in  \mathcal{F}\, |\, 
S\Psi = 0
\right\}.
$$ 
In addition, when $S$ is closed, 
let us denote by $\sigma(S)$ 
the spectrum of a closed operator $S$.  

\begin{definition}
{\rm By {\it ground state energy} 
we mean $\inf\sigma (H_{\mbox{\rm {\tiny QFT}}})$, 
the lowest spectrum of $H_{\mbox{\rm {\tiny QFT}}}$. 
We denote the ground state energy 
by $E_{0}(H_{\mbox{\rm {\tiny QFT}}})$, i.e., 
$E_{0}(H_{\mbox{\rm {\tiny QFT}}}) 
:= \inf\sigma (H_{\mbox{\rm {\tiny QFT}}})$. 
We say {\it $H_{\mbox{\rm {\tiny QFT}}}$ has a 
ground state $\Psi_{\mbox{\rm {\tiny QFT}}}$} 
if $\mathrm{ker}\, ( 
H_{\mbox{\rm {\tiny QFT}}} 
- E_{0}(H_{\mbox{\rm {\tiny QFT}}}))$ 
is not empty and then 
$0 \ne  \Psi_{\mbox{\rm {\tiny QFT}}} 
\in  \mathrm{ker}\, ( 
H_{\mbox{\rm {\tiny QFT}}} 
- E_{0}(H_{\mbox{\rm {\tiny QFT}}}))$. 
We say $\Psi_{\mbox{\rm {\tiny QFT}}}$ 
to be normalized if $\| 
\Psi_{\mbox{\rm {\tiny QFT}}}
\|_{\mathcal{F}} = 1$. 
}
\end{definition}

For simplicity, we set $\widehat{H}_{\mbox{\rm {\tiny QFT}}}$ 
as 
$$
\widehat{H}_{\mbox{\rm {\tiny QFT}}} 
:= H_{\mbox{\rm {\tiny QFT}}} - 
E_{0}(H_{\mbox{\rm {\tiny QFT}}}). 
$$
We always suppose that $\Psi_{\mbox{\rm {\tiny QFT}}}$ has 
been normalized whenever it exists. 

\section{Domain Properties of the Carleman Operator 
for IR Catastrophe}
\label{sec:DPCOIRC}

When the operator-theoretical pull-through (OPPT) 
formula on ground states holds 
in the same way as in \cite{hirokawa-nelson}, 
$a(f)\Psi_{\mbox{\rm {\tiny QFT}}}$ has the expression:    
\begin{eqnarray}
a(f)\Psi_{\mbox{\rm {\tiny QFT}}} 
=  
- \int_{{\mathbb R}^{d}}
\overline{f(k)}
\left( \widehat{H}_{\mbox{\tiny QFT}} 
+ \omega(k)\right)^{-1}
B_{\mbox{\rm {\tiny PT}}}(k)
\Psi_{\mbox{\rm {\tiny QFT}}}
dk     
\label{eq:OPPT} 
\end{eqnarray}
for every $f \in  C_{0}^{\infty}({\mathbb R}^{d}
\setminus \left\{ 0\right\})$. 
This is the operator-theoretical version of 
the symbolical pull-through formula on ground states: 
\begin{eqnarray}
a(k)\Psi_{\mbox{\rm {\tiny QFT}}} 
=  
- \left( \widehat{H}_{\mbox{\tiny QFT}} 
+ \omega(k)\right)^{-1}
B_{\mbox{\rm {\tiny PT}}}(k)
\Psi_{\mbox{\rm {\tiny QFT}}}. 
\label{eq:PT} 
\end{eqnarray}
Then, we have an operator 
$B_{{\mbox{\rm {\tiny PT}}}}(k)$ 
for every $k \in  \mathbb{R}^{d}
\setminus\{ 0\}$ in the integrand of 
Eq.(\ref{eq:OPPT}).  
We can show OPPT formula holds for several models 
in quantum field theory 
\cite{ahh2, hirokawa-nelson, hiroshima}.

In our argument, we assume the following conditions: 
\begin{description}
\item[(Ass.1)] Eq. (\ref{eq:OPPT}) holds and then 
$B_{\mbox{\tiny PT}}(k)$ is determined for 
every $k \in  {\mathbb R}^{d}\setminus 
\left\{ 0\right\}$ as 
an operator acting in ${\mathcal F}$ and then 
$B_{\mbox{\tiny PT}}(\cdot)\Psi$ is measurable for every 
$\Psi \in  D(H_{0})$ 
(i.e., $D(B_{\mbox{\tiny PT}}(k)) \supset D(H_{0})$ 
for every $k \in  \mathbb{R}^{d}\setminus\{ 0\}$ 
and $\left( \Phi\, ,\, 
B_{\mbox{\tiny PT}}(\cdot)\Psi\right)_{\mathcal F} 
: {\mathbb R}^{d} \longrightarrow {\mathbb C}$ is 
measurable for every $\Phi \in  {\mathcal F}$). 
\item[(Ass.2)] $( \widehat{H}_{\mbox{\tiny QFT}} 
+ \omega(k))^{-1}B_{\mbox{\tiny PT}}(k)$ is bounded for every 
$k \in {\mathbb R}^{d}\setminus \left\{ 0\right\}$ and then for 
every $\varepsilon > 0$
\begin{eqnarray}
M_{\varepsilon} := 
\left\{
\int_{|k|>\varepsilon}
\| (\widehat{H}_{\mbox{\rm {\tiny QFT}}} 
+ \omega(k))^{-1}
B_{\mbox{\rm {\tiny PT}}}(k)
\|_{\mathcal{B}(\mathcal{F})}^{2}
dk\right\}^{1/2} < \infty,  
\label{ass:2}
\end{eqnarray}
\end{description}
where $\|\cdot\|_{\mathcal{B}(\mathcal{F})}$ 
denotes the operator norm of $\mathcal{B}(\mathcal{F})$, 
the $C^{*}$-algebra of bounded operators 
on $\mathcal{F}$. 
\hfill\break

For every $\varepsilon \ge 0$, we set 
${\mathbb R}^{d}_{<\varepsilon}$ and 
${\mathbb R}^{d}_{>\varepsilon}$ as 
\begin{eqnarray*}
{\mathbb R}^{d}_{<\varepsilon} 
:= \left\{ k \in {\mathbb R}^{d}\, \big|\, 
|k| < \varepsilon \right\}\quad
\mbox{and}\quad  
{\mathbb R}^{d}_{>\varepsilon} 
:= \left\{ k \in {\mathbb R}^{d}\, \big|\, 
|k| > \varepsilon \right\}, 
\label{eq:devided-space}
\end{eqnarray*}
respectively. 
For every $f \in  L^{2}({\mathbb R}^{d})$ 
we define $f^{<\varepsilon}$ and 
$f^{>\varepsilon}$ in $L^{2}({\mathbb R}^{d})$ 
by 
\begin{eqnarray*}
f^{<\varepsilon}(k) := 
1^{<\varepsilon}(k)f(k)\quad 
\mbox{and}\quad 
f^{>\varepsilon}(k) := 
1^{>\varepsilon}(k)f(k),  
\label{eq:devided-function}
\end{eqnarray*}
where $1^{<\varepsilon}$ 
and $1^{>\varepsilon}$ are 
characteristic functions defined by 
\begin{eqnarray*}
1^{<\varepsilon}(k) := 
\begin{cases}
1 & \text{if $|k| < \varepsilon$}, \\ 
0  & \text{otherwise}, 
\end{cases} 
\quad 
\mbox{and}\quad 
1^{>\varepsilon}(k) := 
\begin{cases}
1 & \text{if $|k| > \varepsilon$}, \\ 
0  & \text{otherwise}. 
\end{cases} 
\end{eqnarray*}
Since we can regard $f^{<\varepsilon}$ and 
$f^{>\varepsilon}$ as functions in 
$L^{2}({\mathbb R}^{d}_{<\varepsilon})$ 
and $L^{2}({\mathbb R}^{d}_{>\varepsilon})$ 
respectively, we often handle them as 
$f^{<\varepsilon} \in 
L^{2}({\mathbb R}^{d}_{<\varepsilon})$ 
and $f^{>\varepsilon} \in 
L^{2}({\mathbb R}^{d}_{>\varepsilon})$ 
in this paper.  

Following this decomposition, 
we introduce some abbreviations: 
\begin{eqnarray*}
&{}& 
d\Gamma_{<\varepsilon}(h^{<\varepsilon}) 
:= d\Gamma_{{\mathbb R}_{<\varepsilon}^{d}}
(h^{<\varepsilon}), 
\qquad 
d\Gamma_{>\varepsilon}(h^{>\varepsilon}) 
:= d\Gamma_{{\mathbb R}_{>\varepsilon}^{d}}
(h^{>\varepsilon}), \\
&{}& 
a^{\sharp}_{<\varepsilon}(f^{<\varepsilon}) 
:= 
a^{\sharp}_{\mathbb{R}^{d}_{<\varepsilon}}
(f^{<\varepsilon}), 
\qquad  
a^{\sharp}_{>\varepsilon}(f^{>\varepsilon}) 
:= 
a^{\sharp}_{\mathbb{R}^{d}_{>\varepsilon}}
(f^{>\varepsilon}),  \\ 
\mbox{and} &{}& 
a^{\sharp}(f) 
:= 
I\otimes 
a^{\sharp}_{\mathrm{b}}(f) 
= 
I\otimes 
a^{\sharp}_{\mathbb{R}^{d}}
(f), 
\end{eqnarray*}
where $a^{\sharp}_{X}$ denotes 
$a_{X}$ or $a^{\dagger}_{X}$.

By Proposition \ref{A-theo-4-55-56}, 
there exists a unitary operator 
$U_{\varepsilon}$ for every $\varepsilon > 0$ 
such that 
\begin{eqnarray*}
&{}& 
U_{\varepsilon}{\mathcal F} 
=  
\mathcal{H}\otimes
\mathcal{F}_{\mathbb{R}^{d}_{<\varepsilon}}
\otimes 
\mathcal{F}_{\mathbb{R}^{d}_{>\varepsilon}} 
\equiv   
\mathcal{H}\otimes\mathcal{F}_{\mathrm{b}}
(L^{2}(\mathbb{R}^{d}_{<\varepsilon}))
\otimes 
\mathcal{F}_{\mathrm{b}}
(L^{2}(\mathbb{R}^{d}_{>\varepsilon})) 
=: \mathcal{F}_{\varepsilon}.  
\label{eq:U1} 
\end{eqnarray*}
Write $U_{\varepsilon}\Psi \in  {\mathcal F}_{\varepsilon}$ 
as $\Psi_{\varepsilon}$ for every $\Psi \in  
{\mathcal F}$, i.e., $\Psi_{\varepsilon} := 
U_{\varepsilon}\Psi$. 
Then, Proposition \ref{A-theo-4-55-56}(ii) 
leads us to the relation:  
\begin{eqnarray} 
&{}& 
U_{\varepsilon}(I\otimes d\Gamma_{\mathrm b}(h))
U_{\varepsilon}^{*} 
= 
\overline{I\otimes d\Gamma_{<\varepsilon}
(h^{<\varepsilon})
\otimes I 
+ 
I\otimes I\otimes d\Gamma_{>\varepsilon}
(h^{>\varepsilon})} 
\label{eq:U2}  
\end{eqnarray}
for every real-valued function $h: 
{\mathbb R}^{d} \to {\mathbb R}$.

We define the \textit{boson number operator} 
$N$ acting in 
${\mathcal F}$ by 
\begin{eqnarray}
N := I\otimes d\Gamma_{\mathrm b}(1).  
\label{eq:def-number-op}
\end{eqnarray}

Symbolically we set the ground-state expectation 
$\langle S\rangle_{\mathrm{gs}}$ for an operator 
$S$ acting in $\mathcal{F}$ by 
$\langle S\rangle_{\mathrm{gs}} 
:= \left( 
\Psi_{\mbox{\rm {\tiny QFT}}}\, ,\, 
S
\Psi_{\mbox{\rm {\tiny QFT}}}
\right)_{\mathcal{F}}$.  
Here we note $\Psi_{\mbox{\rm {\tiny QFT}}}$ 
is normalized, i.e., 
$\|\Psi_{\mbox{\rm {\tiny QFT}}}
\|_{\mathcal{F}} = 1$. 
Then, we can consider $\langle S\rangle_{\mathrm{gs}}$ 
to be finite if $\Psi_{\mbox{\rm {\tiny QFT}}} 
\in  D(S)$, on the other hand, 
to be infinite if $\Psi_{\mbox{\rm {\tiny QFT}}} 
\notin  D(S)$. We note we can write 
$\Psi_{\mbox{\rm {\tiny QFT}}} 
\notin  D(S)$ when $\Psi_{\mbox{\rm {\tiny QFT}}}$ 
does not exist in $\mathcal{F}$. 
That is, 
\begin{eqnarray*}
&{}& 
\langle S\rangle_{\mathrm{gs}} < \infty\,\,\, 
\mbox{if $\Psi_{\mbox{\rm {\tiny QFT}}} 
\in  D(S)$,} 
\label{eq:gs-expectation0'}\\ 
&{}&  
\langle S\rangle_{\mathrm{gs}} = \infty\,\,\, 
\mbox{if $\Psi_{\mbox{\rm {\tiny QFT}}} 
\notin  D(S)$ or $\Psi_{\mbox{\rm {\tiny QFT}}}$ 
does not exist in $\mathcal{F}$}. 
\label{eq:gs-expectation'}
\end{eqnarray*}

\begin{definition} 
\label{def:IRdivergence} 
{\rm We say the {\it infrared (IR) catastrophe occurs} 
if $\Psi_{\mbox{\rm {\tiny QFT}}} \notin 
D(N^{1/2})$ including 
the case where $\Psi_{\mbox{\rm {\tiny QFT}}}$ 
does not exist in $\mathcal{F}$, i.e., 
$\Psi_{\mbox{\rm {\tiny QFT}}} \notin 
\mathcal{F}$.
} 
\end{definition}

\begin{remark}
\label{rem:*}
Since $D(N) \subset D(N^{1/2})$, 
the naive meaning of Definition \ref{def:IRdivergence} 
is symbolically: 
$$ 
\langle N\rangle_{\mathrm{gs}} = 
\left(\Psi_{\mbox{\rm {\tiny QFT}}} \, ,\, 
N\Psi_{\mbox{\rm {\tiny QFT}}}\right)_{\mathcal{F}} 
= \| N^{1/2}\Psi_{\mbox{\rm {\tiny QFT}}} 
\|_{\mathcal{F}}^{2} 
= \infty. 
$$
\end{remark}

For every $\varepsilon > 0$, we 
define $N_{>\varepsilon}$ acting in 
${\mathcal F}$ by 
\begin{eqnarray*}
N_{>\varepsilon} := U_{\varepsilon}^{*}
(I\otimes I\otimes d\Gamma_{>\varepsilon}
(1^{>\varepsilon}))
U_{\varepsilon}. 
\label{eq:number-operator''}
\end{eqnarray*}

\begin{lemma}\cite{gerard-private} 
\label{lemma:gerard-private}
$${\displaystyle D(H_{0}) 
\subset 
\bigcap_{\varepsilon > 0}
D(N_{>\varepsilon}^{1/2}).} 
$$
\end{lemma}

\demo 
Since $1^{>\varepsilon}(k) \le 
(\inf_{|k|>\varepsilon}\omega(k))^{-1}
\omega^{>\varepsilon}(k)$ 
for every $\varepsilon > 0$, 
we have $D(\omega^{>\varepsilon}) \subset  
D(1^{>\varepsilon})$, 
which implies 
$D(I\otimes d\Gamma_{>\varepsilon}
(\omega^{>\varepsilon})) 
\subset 
D(I\otimes d\Gamma_{>\varepsilon}
(1^{>\varepsilon}))$. 
Thus, by Eq.(\ref{eq:U2}) we have  
\begin{eqnarray*}
D(H_{0}) 
&=& D(A\otimes I)\cap D(I\otimes d\Gamma(\omega)) \\ 
&\cong& 
D(A\otimes I\otimes I)
\cap D(I\otimes d\Gamma(\omega^{<\varepsilon})\otimes I)
\cap D(I\otimes I\otimes d\Gamma(\omega^{>\varepsilon})) \\ 
&\subset& 
{\mathcal F}_{\varepsilon}\cap 
D(I\otimes I\otimes d\Gamma(1^{>\varepsilon})) 
= D(I\otimes I\otimes d\Gamma(1^{>\varepsilon})) 
\cong 
D(N_{>\varepsilon}). 
\end{eqnarray*}
Combining the fact that $D(N_{>\varepsilon}) \subset 
D(N_{>\varepsilon}^{1/2})$ with the above 
leads us to our lemma. 
\qed

To find a relation between $N$ and $N_{>\varepsilon}$, 
we introduce the following domain: 
 
\begin{eqnarray*}
D_{\mbox{\tiny CNB}} := 
\biggl\{ \Psi \in  \bigcap_{\varepsilon > 0}
D(N_{>\varepsilon}^{1/2})\, \bigg|\, 
\sup_{\varepsilon > 0}
\| N_{> \varepsilon}^{1/2}\Psi\|_{\mathcal F}^{2} 
< \infty
\biggr\}. 
\label{eq:D1}
\end{eqnarray*}

The following lemma is a mathematical 
establishment of Eq.(\ref{eq:establishment'}). 

\begin{lemma}
\label{lemma:establishment} 
Let $\{f_{\ell}^{>\varepsilon}\}_{\ell=1}^{\infty}$ 
be an arbitrary complete orthonormal system of 
$L^{2}({\mathbb R}^{d}_{>\varepsilon})$ 
for every $\varepsilon >0$. 
Then, 
\begin{eqnarray*}
\| N_{>\varepsilon}^{1/2}\Psi\|_{\mathcal F}^{2} 
= 
\sum_{\ell=0}^{\infty}
\| a(f_{\ell}^{>\varepsilon})
\Psi\|_{\mathcal F}^{2}
= 
\sum_{\ell=0}^{\infty}
\| I\otimes a_{\mathrm{b}}(f_{\ell}^{>\varepsilon})
\Psi\|_{\mathcal F}^{2}
\label{eq:new-5}
\end{eqnarray*}
for every $\Psi \in  D(N_{>\varepsilon}^{1/2})$. 
\end{lemma}

\demo 
By Proposition \ref{RIMS:cor-5.1}, we have
\begin{eqnarray*}
\nonumber 
&{}& 
\| N_{>\varepsilon}^{1/2}\Psi\|_{\mathcal F}^{2} 
= 
\|
I\otimes I\otimes d\Gamma(1_{>\varepsilon})^{1/2}
\Psi_{\varepsilon}
\|_{{\mathcal F}_{\varepsilon}}^{2}  
= 
\sum_{\ell=1}^{\infty}
\| I\otimes I\otimes 
a_{>\varepsilon}(f_{\ell}^{>\varepsilon})
\Psi_{\varepsilon}
\|_{{\mathcal F}_{\varepsilon}}^{2} \\ 
&=& 
\sum_{\ell=1}^{\infty}
\| I\otimes a_{\mathrm{b}}(f_{\ell}^{>\varepsilon})
\Psi\|_{\mathcal F}^{2}.   
\end{eqnarray*}
The above equation, together with 
$\| a(f_{\ell}^{>\varepsilon})
\Psi\|_{\mathcal F}^{2} 
=\| I\otimes a_{\mathrm{b}}(f_{\ell}^{>\varepsilon})
\Psi\|_{\mathcal F}^{2} 
= 
\| I\otimes I\otimes 
a_{>\varepsilon}(f_{\ell}^{>\varepsilon})\Psi_{\varepsilon}
\|_{{\mathcal F}_{\varepsilon}}^{2}$, 
completes the proof of our lemma. 
\qed

The following lemma gives a relation 
between $N$ and $N_{>\varepsilon}$. 
It tells us that for all vectors $\Psi 
\in D(H_{0})$ we can check 
whether $(\Psi\, ,\, N\Psi)_{\mathcal{F}}$ 
converges by taking advantage of 
Lemma \ref{lemma:gerard-private} and estimating 
$\sup_{\varepsilon > 0}\| N_{>\varepsilon}^{1/2}\Psi
\|_{\mathcal F}^{2}$. 
Thus, the following lemma plays an important role to 
prove Theorems \ref{theorem:CNB} and 
\ref{theorem:hougan} below.

\begin{lemma}
\label{lemma:CNB}
$D_{\mbox{\rm {\tiny CNB}}} = D(N^{1/2})$ and 
\begin{eqnarray*}
\sup_{\varepsilon > 0}\| N_{>\varepsilon}^{1/2}\Psi\|_{\mathcal F}^{2} 
= \| N^{1/2}\Psi\|_{\mathcal F}^{2}
\label{eq:CNB'}
\end{eqnarray*}
for $\Psi \in  D_{\mbox{\rm {\tiny CNB}}}$. 
\end{lemma}

\demo 
Let $\Psi \in D(N^{1/2})$ first. 
Then, we can show $\Psi \in \bigcap_{\varepsilon>0}
D(N_{>\varepsilon}^{1/2})$ in the same way as 
the proof of Lemma \ref{lemma:gerard-private}. 
Let $\left\{ f_{\ell}\right\}_{\ell=1}^{\infty}$ 
be an arbitrary complete orthonormal system of 
$L^{2}({\mathbb R}^{d})$. 
We decompose $f_{\ell}$ into 
$f_{\ell}^{<\varepsilon}$ 
and $f_{\ell}^{>\varepsilon}$, i.e., 
$f_{\ell} = f_{\ell}^{<\varepsilon} 
+ f_{\ell}^{>\varepsilon}$. 
Then, evidently 
$\{ f_{\ell}^{>\varepsilon}\}_{\ell=1}^{\infty}$ 
makes a complete orthonormal system of 
$L^{2}({\mathbb R}^{d}_{>\varepsilon})$ and moreover 
$f_{\ell}^{>\varepsilon} \to f_{\ell}$ in 
$L^{2}({\mathbb R}^{d})$ as $\varepsilon \to 0$. 

To obtain 
$\sup_{\varepsilon>0}
\sum_{\ell=1}^{\infty}
\| I\otimes a_{\mathrm{b}}
(f_{\ell}^{>\varepsilon})
\Psi\|_{\mathcal F}^{2} 
= 
\| N^{1/2}\Psi\|_{\mathcal F}^{2}$ 
we carefully revise the method which we used 
in the proof of Proposition \ref{RIMS:cor-5.1}.
For each $M \in  {\mathbb N}$ we can derive  
\begin{eqnarray}
\sum_{\ell=1}^{M}
\| I\otimes a_{\mathrm{b}}(f^{>\varepsilon}_{\ell})
\Psi\|_{\mathcal F}^{2} 
= 
\sum_{n=0}^{\infty}
(n+1)\int_{{\mathbb R}^{dn}}
\Psi_{M,\varepsilon}^{(n)}
(k_{1}, \cdots, k_{n})dk_{1}\cdots dk_{n} 
\label{eq:new-1}
\end{eqnarray}
from the definition of the annihilation operator, 
where we used the representation 
(\ref{eq:V-rep}) and 
\begin{eqnarray*}
\Psi_{M,\varepsilon}^{(n)}
(k_{1}, \cdots, k_{n}) 
:= 
\sum_{\ell=0}^{M} 
\Bigg\| 
\int_{|k|>\varepsilon} 
\overline{f_{\ell}^{>\varepsilon}(k)}\, 
\Psi_{\mathcal H}^{(n+1)}(k, k_{1}, \cdots, k_{n})dk
\Bigg\|_{\mathcal H}^{2}. 
\label{eq:new-2}
\end{eqnarray*}
Let $\{ e_{p}\}_{p=1}^{\infty}$ be an arbitrary 
complete orthonormal system of ${\mathcal H}$. 
Then, we have 
\begin{eqnarray*}
\Psi_{M,\varepsilon}^{(n)}
(k_{1}, \cdots, k_{n}) 
=   
\sum_{\ell=1}^{M}\sum_{p=1}^{\infty}
\Bigg| 
\int_{{\mathbb R}_{>\varepsilon}^{d}}
\left(f_{\ell}^{>\varepsilon}(k) e_{p}\, ,\, 
\Psi_{\mathcal H}^{(n+1)}(k, k_{1}, \cdots, k_{n})
\right)_{\mathcal H}dk
\Bigg|^{2}. 
\end{eqnarray*}
We note here that we can regard 
$\Psi_{\mathcal H}^{(n+1)}(\cdot, k_{1}, \cdots, k_{n})$ 
as a function in $L^{2}({\mathbb R}_{>\varepsilon}^{d}; 
{\mathcal H})$ for a.e. $(k_{1}, \cdots, k_{n})$ 
because $\Psi_{\mathcal H}^{(n+1)}(\cdot, k_{1}, \cdots, k_{n}) 
\in L^{2}({\mathbb R}^{d}; {\mathcal H})$ 
for a.e. $(k_{1}, \cdots, k_{n})$. 
Moreover, $\{ f_{\ell}^{>\varepsilon}e_{p}\}_{\ell,p=1}^{\infty}$ 
makes a complete orthonormal system of 
$L^{2}({\mathbb R}_{>\varepsilon}^{d}; {\mathcal H})$. 
Hence it follows from Bessel's inequality that 
\begin{eqnarray*}
\Psi_{M,\varepsilon}^{(n)}
(k_{1}, \cdots, k_{n}) 
&\le& 
\|\Psi^{(n+1)}_{\mathcal H}
(\cdot, k_{1}, \cdots, k_{n})\|_{
L^{2}({\mathbb R}_{>\varepsilon}^{d}; 
{\mathcal H})
}^{2} \\  
&\le& 
\|\Psi^{(n+1)}_{\mathcal H}
(\cdot, k_{1}, \cdots, k_{n})\|_{
L^{2}({\mathbb R}^{d}; {\mathcal H})
}^{2}.  
\end{eqnarray*} 
Thus, we have 
\begin{eqnarray}
\Psi_{M,\varepsilon}^{(n)}
(k_{1}, \cdots, k_{n}) 
\nearrow 
\int_{|k|>\varepsilon} 
\|\Psi^{(n+1)}_{\mathcal H}(k,k_{1},\cdots,k_{n})
\|_{\mathcal H}^{2}dk
\label{eq:new-3}
\end{eqnarray}
as $M\to\infty$. 
Applying Lebesgue's monotone convergence theorem and 
Fubini's theorem to Eqs.(\ref{eq:new-1}) and (\ref{eq:new-3}), 
we reach the expression:  
\begin{eqnarray*}
&{}& 
\sum_{\ell=1}^{\infty}
\| I\otimes a_{\mathrm{b}}(f_{\ell}^{>\varepsilon})
\Psi\|_{\mathcal F}^{2} \\ 
&=& 
\sum_{n=0}^{\infty}(n+1)
\int_{|k|>\varepsilon} 
\int_{{\mathbb R}^{dn}}
\|\Psi^{(n+1)}(k,k_{1},\cdots,k_{n})\|_{\mathcal H}^{2}
dk_{1}\cdots dk_{n}dk.
\end{eqnarray*}
From this expression, we know that 
$\sum_{\ell=1}^{\infty}
\| I\otimes a_{\mathrm{b}}
(f_{\ell}^{>\varepsilon})
\Psi\|_{\mathcal F}^{2}$ 
is increasing as $\varepsilon \to 0$. 
So, applying Lebesgue's monotone 
convergence theorem to the above 
equation yields 
\begin{eqnarray}
\nonumber 
&{}& 
\sup_{\varepsilon>0}
\sum_{\ell=1}^{\infty}
\| I\otimes a_{\mathrm{b}}
(f_{\ell}^{>\varepsilon})
\Psi\|_{\mathcal F}^{2} 
= 
\lim_{\varepsilon\to 0}
\sum_{\ell=1}^{\infty}
\| I\otimes a_{\mathrm{b}}
(f_{\ell}^{>\varepsilon})
\Psi\|_{\mathcal F}^{2} \\ 
\nonumber 
&=& 
\sum_{n=0}^{\infty}(n+1)
\int_{{\mathbb R}^{d(n+1)}}
\|\Psi^{(n+1)}(k_{1},\cdots,k_{n+1})\|_{\mathcal H}^{2}
dk_{1}\cdots dk_{n+1}  \\ 
&=& 
\| N^{1/2}\Psi_{\mathcal H}\|_{
\bigoplus_{n}L_{\mathrm {sym}}^{2}
({\mathbb R}^{dn};{\mathcal H})}^{2} 
= 
\| N^{1/2}\Psi\|_{\mathcal F}^{2}.  
\label{eq:new-4}
\end{eqnarray}
By Lemma \ref{lemma:establishment} and 
Eq. (\ref{eq:new-4}), 
we obtain that 
\begin{eqnarray}
\sup_{\varepsilon>0}
\| N_{>\varepsilon}^{1/2}\Psi\|_{\mathcal F}^{2} 
= 
\| N^{1/2}\Psi\|_{\mathcal F}^{2} < \infty,  
\label{eq:3-12-0}
\end{eqnarray}
which means $\Psi \in  D_{\mbox{\rm {\tiny CNB}}}$. 

Conversely, let $\Psi = \oplus\sum_{n=0}^{\infty}
\Psi^{(n)} \in D_{\mbox{\rm {\tiny CNB}}}$. 
Using symmetry of $\Psi^{(n)}(k_{1},\cdots, k_{n})$, 
we have  
\begin{eqnarray*}
&{}& 
\| N_{>\varepsilon}^{1/2}\Psi\|_{\mathcal F}^{2} 
= 
\sum_{n=1}^{\infty}
\| I\otimes I\otimes d\Gamma_{>\varepsilon}
(1_{>\varepsilon})^{1/2}
\Psi_{\varepsilon}^{(n)}
\|_{\mathcal{F}_{\varepsilon}}^{2} \\ 
&=& 
\sum_{n=1}^{\infty}\Biggr\{ 
n \int_{|k_{1}|, \cdots, |k_{n}| > \varepsilon}
\| \Psi^{(n)}(k_{1},\cdots, k_{n})\|_{\mathcal H}^{2}
dk_{1}\cdots dk_{n} \\ 
&{}&  
+ 
\sum_{j=1}^{n-1}
\begin{pmatrix}
n \\ 
j 
\end{pmatrix} 
j \int_{|k_{1}|, \cdots, |k_{j}| > \varepsilon\, ;\, 
|k_{j+1}|, \cdots, |k_{n}| < \varepsilon}
\| \Psi^{(n)}(k_{1},\cdots, k_{n})\|_{\mathcal H}^{2}
dk_{1}\cdots dk_{n} \\ 
&{}&  
+ 
\int_{|k_{1}|, \cdots, |k_{n}| < \varepsilon}
\| \Psi^{(n)}(k_{1},\cdots, k_{n})\|_{\mathcal H}^{2}
dk_{1}\cdots dk_{n} 
\Biggl\}\\ 
&\ge& 
\sum_{n=1}^{\infty}
n \int_{|k_{1}|, \cdots, |k_{n}| > \varepsilon}
\| \Psi^{(n)}(k_{1},\cdots, k_{n})\|_{\mathcal H}^{2}
dk_{1}\cdots dk_{n} 
=: \Theta(\varepsilon). 
\end{eqnarray*}
Here let us set $A_{\nu}(\varepsilon)$ as 
$$A_{\nu}(\varepsilon) := 
\sum_{n=1}^{\nu}
n \int_{|k_{1}|, \cdots, |k_{n}| > \varepsilon}
\| \Psi^{(n)}(k_{1},\cdots, k_{n})\|_{\mathcal H}^{2}
dk_{1}\cdots dk_{n}.$$
Then, Lebesgue's monotone convergence theorem 
implies  
\begin{eqnarray}
A_{\nu}(\varepsilon) 
\longrightarrow 
\sum_{n=1}^{\nu}
n \int_{{\mathbb R}^{dn}}
\| \Psi^{(n)}(k_{1},\cdots, k_{n})\|_{\mathcal H}^{2}
dk_{1}\cdots dk_{n} 
=: A_{\nu}
\label{eq:def-Anu}
\end{eqnarray}
as $\varepsilon\to 0$. 
Since $\Theta(\varepsilon)$ is increasing as 
$\varepsilon$ tends to $0$, 
we have 
\begin{eqnarray} 
\infty > 
\sup_{\varepsilon > 0}
\| N_{>\varepsilon}^{1/2}\Psi\|_{\mathcal F}^{2} 
\ge 
\lim_{\varepsilon\to 0}
\Theta(\varepsilon) 
\ge 
\lim_{\varepsilon\to 0}
A_{\nu}(\varepsilon) = A_{\nu}
\label{eq:3-12-1}
\end{eqnarray}
for each $\nu \in  {\mathbb N}$. 
By Eqs.(\ref{eq:def-Anu}) and (\ref{eq:3-12-1}), 
$\left\{ A_{\nu}\right\}_{\nu \in 
{\mathbb N}}$ is monotone increasing and bounded. 
Therefore, $\lim_{\nu\to\infty}A_{\nu}$ 
exists and then we have 
$$
\infty > \lim_{\nu\to\infty}A_{\nu} 
= 
\sum_{n=1}^{\infty}
n \int_{{\mathbb R}^{dn}}
\| \Psi^{(n)}(k_{1},\cdots, k_{n})\|_{\mathcal H}^{2}
dk_{1}\cdots dk_{n} 
= \| N^{1/2}\Psi\|_{\mathcal F}^{2}.   
$$ 
Hence it follows from this and Eq.(\ref{eq:3-12-1}) 
that 
\begin{eqnarray}
\infty > \sup_{\varepsilon > 0}
\| N_{>\varepsilon}^{1/2}\Psi\|_{\mathcal F}^{2} 
\ge 
\| N^{1/2}\Psi\|_{\mathcal F}^{2}
\label{eq:3-12-2}. 
\end{eqnarray}
So, we reach the conclusion that 
$\Psi \in  D(N^{1/2})$, and thus, 
Eq.(\ref{eq:3-12-0}) holds. 
\qed

In \cite{hirokawa-mecha}, 
the author tried to use Fatou's lemma to 
prove Eq.(\ref{eq:3-12-2}). 
But there was a mistake in his proof. 
In \cite{br2}, Bruneau completes the author's idea. 
Namely, we have 
\begin{eqnarray*}  
\| N_{>\varepsilon}^{1/2}\|_{\mathcal F}^{2} 
= 
\sum_{n=0}^{\infty} 
\sum_{j=0}^{n}
\begin{pmatrix}
n \\ j
\end{pmatrix}
j \|\underbrace{(\overline{p}_{\varepsilon}\otimes 
\cdots\otimes \overline{p}_{\varepsilon})}_{n-j}
\otimes \underbrace{(p_{\varepsilon}\otimes 
\cdots\otimes p_{\varepsilon})}_{j}
\Psi^{(n)}\|_{\mathcal F}^{2},
\label{eq:bruneau}
\end{eqnarray*}
where $p_{\varepsilon}$ is the orthogonal 
projection from 
$L^{2}({\mathbb R}^{d})$ onto 
$L^{2}({\mathbb R}^{d}_{>\varepsilon})$ and 
$\overline{p}_{\varepsilon} := 1 - p_{\varepsilon}$. 
So, applying Fatou's lemma, we have also 
Eq.(\ref{eq:3-12-2}).

\begin{definition}
\label{def:CarlemanOp}
{\rm 
When a ground state $\Psi_{\mbox{\rm {\tiny QFT}}}$ 
of $H_{\mbox{\rm {\tiny QFT}}}$ exists, 
we can define an ${\mathcal F}$-valued function 
$K_{\mbox{\tiny PT}} : {\mathbb R}^{d}\setminus 
\left\{ 0\right\} \longrightarrow 
{\mathcal F}$ by    
\begin{eqnarray} 
K_{\mbox{\rm {\tiny PT}}}(k) :=  
(\widehat{H}_{\mbox{\rm {\tiny QFT}}} 
+ \omega(k))^{-1}
B_{\mbox{\rm {\tiny PT}}}(k)\Psi_{\mbox{\rm {\tiny QFT}}} 
\label{eq:Carleman-kernel} 
\end{eqnarray}
for every $k \in  {\mathbb R}^{d}
\setminus \{ 0\}$ 
since $( \widehat{H}_{\mbox{\tiny QFT}} 
+ \omega(k))^{-1}B_{\mbox{\tiny PT}}(k)$ 
is a bounded operator on $\mathcal{F}$ 
for every $k \in {\mathbb R}^{d}\setminus 
\left\{ 0\right\}$ by (Ass.2). 
$K_{\mbox{\tiny PT}}$ defined by 
Eq.(\ref{eq:Carleman-kernel}) 
is measurable by (Ass.1). 
For the ground state $\Psi_{\mbox{\tiny QFT}}$,  
we define the {\it maximal Carleman operator} 
$T_{\mbox{\tiny PT}} : 
{\mathcal F} \to L^{2}({\mathbb R}^{d})$ 
induced by $K_{\mbox{\rm {\tiny PT}}}$ 
in the following:   
\begin{eqnarray} 
&{}& 
D(T_{\mbox{\rm {\tiny PT}}}) := 
\left\{ \Phi \in {\mathcal F}\, \Big|\, 
\left( K_{\mbox{\rm {\tiny PT}}}(\cdot)\, ,\, 
\Phi\right)_{\mathcal F} \in  L^{2}({\mathbb R}^{d})\right\}, 
\label{eq:Carleman-domain} \\ 
\nonumber 
&{}& 
\left( T_{\mbox{\rm {\tiny PT}}}\Phi\right)(k)  
:= \left( K_{\mbox{\rm {\tiny PT}}}(k)
\, ,\, \Phi\right)_{\mathcal F},
\,\,\, \Phi \in  D(T_{\mbox{\tiny PT}}), 
\label{eq:Carleman-action} 
\end{eqnarray} 
for every $k \in {\mathbb R}^{d}
\setminus\{ 0\}$. 
Then, we call $K_{\mbox{\rm {\tiny PT}}}$ 
the {\it inducing function} of 
$T_{\mbox{\rm {\tiny PT}}}$. 
When $K_{\mbox{\rm {\tiny PT}}}$ has 
a singularity at $k = 0$, 
we call it \textit{IR singularity 
of $T_{\mbox{\rm {\tiny PT}}}$}.  
}
\end{definition}
We note that $T_{\mbox{\rm {\tiny PT}}}$ 
is closed by \cite[Theorem 6.13]{weidmann}.

The following theorem is stated in 
\cite[Theorem 2.9]{hiroshima} by Hiroshima. 
We now give it another proof 
by obeying the Derezi\'{n}ski-G\'{e}rard idea 
\cite{gerard-private} 
and taking advantage of Lemma \ref{lemma:CNB}.

\begin{theorem}
\label{theorem:CNB}
Assume {\rm $D(\mbox{\rm $H_{\mbox{\tiny QFT}}$}) = 
D(H_{0})$} and there exists a ground state {\rm 
$\Psi_{\mbox{\tiny QFT}}$} of {\rm $H_{\mbox{\tiny QFT}}$}. 
Then, the following conditions are equivalent: 
\begin{description}
\item[(i)] $\Psi_{\mbox{\rm {\tiny QFT}}} 
\in  D(N^{1/2})$. 
\item[(ii)] $\| K_{\mbox{\rm {\tiny PT}}}(\cdot)
\|_{\mathcal F} \in  L^{2}({\mathbb R}^{d})$.  
\item[(iii)] $T_{\mbox{\rm {\rm {\tiny PT}}}}$ is a 
Hilbert-Schmidt operator.
\end{description}  
If one of them holds, then 
$\| T_{\mbox{\rm {\rm {\tiny PT}}}}\| 
\le M_{0} := \lim_{\varepsilon\to 0}
M_{\varepsilon} <\infty$.
\end{theorem}

\demo  
Before proving this theorem, 
we first show Eq.(\ref{eq:clueCNB}) below. 
Let $\left\{f_{\ell}^{>\varepsilon}\right\}_{\ell=1}^{\infty}$ 
be an arbitrary complete orthonormal system 
of $L^{2}({\mathbb R}^{d}_{>\varepsilon})$. 
Lemma \ref{lemma:gerard-private} leads us to the fact that 
$\Psi_{\mbox{\tiny QFT}} \in \bigcap_{\varepsilon >0} 
D(N_{>\varepsilon}^{1/2})$. 
By Lemma \ref{lemma:establishment}, we have 
\begin{eqnarray}
\infty > \| N_{>\varepsilon}^{1/2}
\mbox{\rm $\Psi_{\mbox{\tiny QFT}}$}\|_{\mathcal F}^{2} 
= \sum_{\ell=1}^{\infty}\| a(f_{\ell}^{>\varepsilon})
\mbox{\rm $\Psi_{\mbox{\tiny QFT}}$}
\|_{\mathcal F}^{2},
\label{eq:3-12-4}
\end{eqnarray}
Here, in the same way as for 
$T_{\mbox{\rm {\tiny PT}}}$,  
we define the maximal Carleman operator 
$T_{\varepsilon} : 
{\mathcal F} \to L^{2}({\mathbb R}^{d}_{>\varepsilon})$ 
by  
\begin{eqnarray} 
\nonumber 
&{}& 
D(T_{\varepsilon}) := 
\left\{ \Phi \in {\mathcal F}\, \Big|\, 
\left( K_{\mbox{\tiny PT}}(\cdot)\, ,\, 
\Phi\right)_{\mathcal F} \in  
L^{2}({\mathbb R}^{d}_{>\varepsilon})\right\}, 
\label{eq:Carleman-domain-epsilon} \\ 
&{}& 
\left( T_{\varepsilon}\Phi\right)(k)  
:= \left( K_{\mbox{\tiny PT}}(k)\, ,\, \Phi\right)_{\mathcal F},
\,\,\, \Phi \in  D(T_{\varepsilon}),  
\label{eq:Carleman-action-epsilon} 
\end{eqnarray}
for every $k \in {\mathbb R}^{d}_{>\varepsilon}$. 
By the condition (\ref{ass:2}), 
$T_{\varepsilon}$ is a bounded operator with 
$D(T_{\varepsilon}) = {\mathcal F}$ and  
$\| T_{\varepsilon}\| \le M_{\varepsilon}$. 
Thus, the adjoint operator $T_{\varepsilon}^{*}$ of 
$T_{\varepsilon}$ is well-defined. 
Here, remember that we employed the normalized ground state 
$\Psi_{\mbox{\rm {\tiny QFT}}}$ if it exists. 
By Eqs.(\ref{eq:OPPT}), (\ref{eq:Carleman-kernel}), 
and (\ref{eq:Carleman-action-epsilon}), 
we have 
\begin{eqnarray*}
\left(\Phi\, ,\, I\otimes a(f_{\ell}^{>\varepsilon})
\Psi_{\mbox{\rm {\tiny QFT}}}\right)_{\mathcal F}
= -\, 
\overline{
\left( \overline{f_{\ell}^{>\varepsilon}}\, ,\, 
T_{\varepsilon}\Phi\right)_{\mathcal F}
} 
= 
-\, 
\left( \Phi\, ,\, T_{\varepsilon}^{*}
\overline{f_{\ell}^{>\varepsilon}}
\right)_{\mathcal F}
\end{eqnarray*}
for every $\Phi \in {\mathcal F}$, 
which implies 
\begin{eqnarray}
a(f_{\ell}^{>\varepsilon})
\Psi_{\mbox{\rm {\tiny QFT}}} 
= -\, T_{\varepsilon}^{*}
\overline{f_{\ell}^{>\varepsilon}} 
\label{eq:Carleman-adjoint}
\end{eqnarray}
for each $\ell \in  {\mathbb N}$. 
Thus, by applying \cite[Theorems VI.18]{rs1} 
to Eqs.(\ref{eq:3-12-4}) and (\ref{eq:Carleman-adjoint}),  
we have 
\begin{eqnarray*}  
\infty > \| N_{>\varepsilon}^{1/2}
\Psi_{\mbox{\rm {\tiny QFT}}}\|_{\mathcal F}^{2}  
&=&   
\sum_{\nu=0}^{\infty}
\| T_{\varepsilon}^{*}\overline{f}_{\nu}\|_{\mathcal F}^{2}  
= 
\sum_{\nu=0}^{\infty}
\left( \overline{f}_{\nu}\, ,\, 
T_{\varepsilon}T_{\varepsilon}^{*}
\overline{f}_{\nu}\right)_{L^{2}
({\mathbb R}^{d}_{>\varepsilon})}   
=   
{\rm tr}(T_{\varepsilon}T_{\varepsilon}^{*}). 
\end{eqnarray*} 
Thus, \cite[VI.22]{rs1} tells us that 
$T_{\varepsilon}^{*}$ is a Hilbert-Schmidt operator 
and thus $T_{\varepsilon}$ is also a Hilbert-Schmidt operator.
Thus, \cite[Theorems 6.12, 6.13]{weidmann} leads us 
to the fact that 
$\|K_{\mbox{\rm {\tiny PT}}}(\cdot)\|_{\mathcal F} 
\in  L^{2}({\mathbb R}_{>\varepsilon}^{d})$. 
Let $\{\Phi_{p}\}_{p=1}^{\infty}$ be an arbitrary complete 
orthonormal system of ${\mathcal F}$. 
Then, by Lemma \ref{lemma:establishment}, and 
Eqs.(\ref{eq:OPPT}) and (\ref{eq:Carleman-kernel} ), 
we have 
\begin{eqnarray*}
&{}&
\| N_{>\varepsilon}^{1/2}\Psi_{\mbox{\rm {\tiny PT}}}
\|_{\mathcal F}^{2} 
= 
\sum_{\ell=1}^{\infty} 
\| I\otimes a(f_{\ell}^{>\varepsilon})
\Psi_{\mbox{\rm {\tiny PT}}}
\|_{\mathcal F}^{2} 
= 
\sum_{\ell=1}^{\infty} 
\Big\| 
\left( f_{\ell}^{>\varepsilon}\, , \, 
K_{\mbox{\rm {\tiny PT}}}
\right)_{L^{2}({\mathbb R}_{>\varepsilon}^{d})}
\Big\|_{\mathcal F}^{2} \\  
&=& 
\sum_{\ell=1}^{\infty}\sum_{p=1}^{\infty} 
\Bigg| 
\int_{|k|>\varepsilon}
\left( f_{\ell}^{>\varepsilon}(k)\Phi_{p}\, ,\, 
K_{\mbox{\rm {\rm {\tiny PT}}}}(k)dk
\right)_{\mathcal F}
\Bigg|^{2}.  
\end{eqnarray*}
Since $\{ f_{\ell}^{>\varepsilon}
\Phi_{p}\}_{\ell,p=0}^{\infty}$ 
makes a complete orthonormal system of 
$L^{2}({\mathbb R}_{>\varepsilon}^{d};{\mathcal F})$, 
this equation yields that $\| N_{>\varepsilon}^{1/2}
\Psi_{\mbox{\rm {\tiny QFT}}}\|_{\mathcal F}^{2} 
= \| K_{\mbox{\rm {\tiny PT}}}(\cdot)\|_{
L^{2}({\mathbb R}_{>\varepsilon}^{d};{\mathcal F})}^{2}$. 
Namely, 
\begin{eqnarray}
\| N_{>\varepsilon}^{1/2}
\Psi_{\mbox{\rm {\tiny QFT}}}\|_{\mathcal F}^{2} 
= 
\int_{|k|>\varepsilon}
\| K_{\mbox{\rm {\tiny PT}}}(k)\|_{\mathcal F}^{2}dk 
< \infty. 
\label{eq:clueCNB}
\end{eqnarray}

We start by showing the equivalence of (i) and (ii). 
It follows immediately from Lebesgue's monotone convergence 
theorem, Lemma \ref{lemma:CNB}, and 
Eq.(\ref{eq:clueCNB}).  

We proceed to the the equivalence of (ii) and (iii). 
It follows directly from \cite[Theorem 6.12]{weidmann} 
that (iii) implies (ii). 
Thus, conversely, we assume 
$\| K_{\mbox{\rm {\tiny PT}}}(\cdot)\|_{\mathcal F} 
\in  L^{2}({\mathbb R}^{d})$ now. 
Then, for every $\Phi \in  {\mathcal F}$ 
we have 
$$
\int_{{\mathbb R}^{d}}
\big|\left(
K_{\mbox{\rm {\tiny PT}}}(k)\, ,\, \Phi
\right)_{\mathcal F}\big|^{2}dk 
\le 
\int_{{\mathbb R}^{d}}
\| K_{\mbox{\rm {\tiny PT}}}(k)\|_{\mathcal F}^{2}dk\, 
\|\Phi\|_{\mathcal F}^{2} 
< \infty 
$$
by Schwarz's inequality, 
which implies that $T_{\mbox{\rm {\tiny PT}}}$ is 
a bounded operator with $D(T_{\mbox{\rm {\tiny PT}}}) 
= {\mathcal F}$ and $\|T_{\mbox{\rm {\tiny PT}}}\| 
\le M_{0}$. 
Obeying \cite[Theorem 6.12]{weidmann}, 
$T_{\mbox{\rm {\tiny PT}}}$ is the restriction of a 
Hilbert-Schmidt operator since 
$\| K_{\mbox{\rm {\tiny PT}}}(\cdot)\|_{\mathcal F} 
\in  L^{2}({\mathbb R}^{d})$. 
Therefore, $T_{\mbox{\rm {\tiny PT}}}$ itself 
is a Hilbert-Schmidt operator. 
\qed

Theorem \ref{theorem:CNB} tells us that 
$D(T_{\mbox{\rm {\tiny PT}}}) = {\mathcal F}$ 
if $\Psi_{\mbox{\rm {\tiny QFT}}}$ exists in 
$D(N^{1/2})$. 
Thus, it is trivial that $D(N^{1/2}) \subset 
D(T_{\mbox{\rm {\tiny PT}}})$ in this case. 
But, more generally, we have this relation 
in the following theorem, 
even though $\Psi_{\mbox{\rm {\tiny QFT}}}$ exists 
outside $D(N^{1/2})$. 
This theorem is proved by 
taking advantage of Lemma \ref{lemma:CNB}:

\begin{theorem}
\label{theorem:hougan}
Suppose that $D(H_{\mbox{\rm {\tiny QFT}}}) 
= D(H_{0})$. 
If a ground state $\Psi_{\mbox{\rm {\tiny QFT}}}$ 
of $H_{\mbox{\rm {\tiny QFT}}}$ exists, 
then 
\begin{eqnarray*}
D(T_{\mbox{\rm {\tiny PT}}}) \supset 
D(N^{1/2}). 
\label{eq:hougan}
\end{eqnarray*}
\end{theorem}

\demo  
Let $\Phi \in  D(N^{1/2})$. 
Then, $\Phi \in  D_{\mbox{\tiny CNB}}$ 
by Lemma \ref{lemma:CNB}.  
We define a functional  $F_{\Phi}: 
C_{0}^{\infty}({\mathbb R}^{d}_{>\varepsilon}) 
\to {\mathbb C}$ by
$F_{\Phi}(f) := 
\left( a(f)\Psi_{\mbox{\tiny QFT}}\, ,\, 
\Phi\right)_{\mathcal F}$ for every 
$f \in  C_{0}^{\infty}({\mathbb R}^{d}_{>\varepsilon})$. 
Since $\Phi \in  D(N_{>\varepsilon}^{1/2})$ 
for every $\varepsilon > 0$ and 
$\Psi_{\mbox{\tiny QFT}}$ is normalized, 
Proposition \ref{proposition:2nd-quantization} 
leads us to the inequality: 
$$| F_{\Phi}(f)| 
\le 
\| f\|_{L^{2}({\mathbb R}^{d}_{>\varepsilon})} 
\| (N_{>\varepsilon} + I)^{1/2}\Phi\|_{\mathcal F}.$$ 
Since $C_{0}^{\infty}({\mathbb R}^{d}_{>\varepsilon})$ 
is dense in $L^{2}({\mathbb R}^{d}_{>\varepsilon})$, 
we have a bounded functional 
from $L^{2}({\mathbb R}^{d}_{>\varepsilon})$ 
to ${\mathbb C}$ as the extension of $F_{\Phi}$. 
We denote it by the same symbol, 
i.e., $F_{\Phi}(f) : L^{2}({\mathbb R}^{d}_{>\varepsilon}) 
\to {\mathbb C}$.  
By Riesz's lemma, there exists $u_{\Phi} 
\in  L^{2}({\mathbb R}^{d}_{>\varepsilon})$ 
such that 
\begin{eqnarray}
\nonumber 
&{}& 
F_{\Phi}(f) = \left( u_{\Phi}\, ,\, 
f\right)_{L^{2}({\mathbb R}^{d}_{>\varepsilon})}, 
\label{eq:2/6-1} \\ 
&{}&  
\| u_{\Phi} \|_{L^{2}({\mathbb R}^{d}_{>\varepsilon})} 
= \| F_{\Phi}\|_{L^{2}({\mathbb R}^{d}_{>\varepsilon})^{*}} 
\le \| (N_{>\varepsilon} + I)^{1/2}\Phi\|_{\mathcal F},  
\label{eq:2/6-2}
\end{eqnarray}
where $L^{2}({\mathbb R}^{d}_{>\varepsilon})^{*}$ 
is the dual space of $L^{2}({\mathbb R}^{d}_{>\varepsilon})$. 
It follows from (Ass.1) that for $f \in  C_{0}^{\infty}
({\mathbb R}^{d}_{>\varepsilon})$ 
\begin{eqnarray*}
&{}& 
\left( f\, ,\, u_{\Phi}
\right)_{L^{2}({\mathbb R}^{d}_{>\varepsilon})} 
= \overline{F_{\Phi}(f)} 
= \left( \Phi\, ,\, a(f)\Psi_{\mbox{\rm {\tiny QFT}}}
\right)_{\mathcal F} \\ 
&=& 
- \int_{|k|>\varepsilon}\overline{f(k)}
\left( \Phi\, ,\, (\widehat{H}_{\mbox{\rm {\tiny QFT}}} 
+ \omega(k))^{-1}B_{\mbox{\rm {\tiny PT}}}(k)
\Psi_{\mbox{\rm {\tiny QFT}}}\right)_{\mathcal F}dk. 
\end{eqnarray*}
Since $C_{0}^{\infty}
({\mathbb R}^{d}_{>\varepsilon})$ is dense in 
$L^{2}({\mathbb R}^{d}_{>\varepsilon})$, 
$$L^{2}({\mathbb R}^{d}_{>\varepsilon}) 
\ni u_{\Phi} = 
- \left( \Phi\, ,\, (\widehat{H}_{\mbox{\rm {\tiny QFT}}} 
+ \omega(\cdot))^{-1}B_{\mbox{\rm {\tiny PT}}}(\cdot)
\Psi_{\mbox{\rm {\tiny QFT}}}\right)_{\mathcal F}. 
$$
By this and the inequality 
(\ref{eq:2/6-2}), we have  
\begin{eqnarray*} 
&{}& \int_{|k|>\varepsilon} 
\biggl|\left( \Phi\, ,\, 
(\widehat{H}_{\mbox{\rm {\tiny QFT}}} 
+ \omega(k))^{-1}B_{\mbox{\rm {\tiny PT}}}(k)
\Psi_{\mbox{\rm {\tiny QFT}}}\right)_{\mathcal F}
\biggr|^{2}dk \\ 
&=& \| u_{\Phi}\|_{L^{2}
({\mathbb R}^{d}_{>\varepsilon})}^{2} 
\le  
\| (N_{>\varepsilon} + I)^{1/2}
\Phi\|_{\mathcal F}^{2}  \\ 
&\le&
 \sup_{\varepsilon > 0}\| N_{>\varepsilon}^{1/2}
\Phi\|_{\mathcal F}^{1/2} 
+ \| \Phi\|_{\mathcal F}^{1/2} < \infty, 
\end{eqnarray*}
since $\Phi \in  D_{\mbox{\tiny CNB}}$. 
So, Lebesgue's monotone convergence theorem 
gives the following estimate:  
\begin{eqnarray*}
&{}& 
\int_{{\mathbb R}^{d}} 
\biggl|\left( \Phi\, ,\, 
(\widehat{H}_{\mbox{\rm {\tiny QFT}}} 
+ \omega(k))^{-1}B_{\mbox{\rm {\tiny PT}}}(k)
\Psi_{\mbox{\rm {\tiny QFT}}}\right)_{\mathcal F}
\biggr|^{2}dk \\ 
&=&  
\lim_{\varepsilon\to 0}
\int_{|k|>\varepsilon} 
\biggl|\left( \Phi\, ,\, (\widehat{H}_{\mbox{\rm {\tiny QFT}}} 
+ \omega(k))^{-1}B_{\mbox{\rm {\tiny PT}}}(k)
\Psi_{\mbox{\rm {\tiny QFT}}}\right)_{\mathcal F}
\biggr|^{2}dk   \\ 
&<& \sup_{\varepsilon > 0}\| N_{>\varepsilon}^{1/2}
\Phi\|_{\mathcal F}^{1/2} 
+ \| \Phi\|_{\mathcal F}^{1/2} < \infty. 
\end{eqnarray*}
Thus, $\left( K_{\mbox{\rm {\tiny PT}}}(\cdot)\, ,\, 
\Phi\right) \in  L^{2}({\mathbb R}^{d})$, 
which means $\Phi \in  D(T_{\mbox{\tiny PT}})$. 
\qed

Theorem \ref{theorem:hougan} gives a sufficient condition 
of IR catastrophe:  

\begin{corollary}
\label{crl:IR-catastrophe}
Suppose that $D(H_{\mbox{{\rm \tiny QFT}}}) = D(H_{0})$. 
If $\Psi_{\mbox{\rm {\tiny QFT}}} \notin  
D(T_{\mbox{\rm {\tiny PT}}})$, 
then IR catastrophe occurs. 
\end{corollary}

Thus, next problem is when $\Psi_{\mbox{\rm {\tiny QFT}}}$ 
is not in $D(T_{\mbox{\rm {\tiny PT}}})$ 
if $\Psi_{\mbox{\rm {\tiny QFT}}}$ exists. 
Theorem \ref{corollary:absence2-2} below deals with 
this question in the case where IR singularity of 
$T_{\mbox{\rm {\tiny PT}}}$ is determined 
by the singularity of a function $\lambda$ 
on $\mathbb{R}^{d}$. 
To prove Theorem \ref{corollary:absence2-2}, 
we need the following property of the domain of the 
Carleman operator:

\begin{theorem}
\label{lemma:zero} 
Suppose $D(H_{\mbox{\rm {\tiny QFT}}}) = D(H_{0})$. 
Assume a function $\lambda$ on $\mathbb{R}^{d}$ 
represents IR singularity of $T_{\mbox{\rm {\tiny PT}}}$ 
as the following (1)--(3): 
\begin{description}
\item[(1)] there is an $\varepsilon_{0} > 0$ 
such that $\lambda(k) \ne 0$ for every $k \in  \mathbb{R}^{d}$ 
with $0 < |k| < \varepsilon_{0}$, 
\item[(2)] $\lambda/\omega \notin  L^{2}(K)$ 
for every neighborhood $K$ of $k=0$, 
\item[(3)] there is an operator $B_{0}(0)$ acting in 
$\mathcal{F}$ such that $\lambda(k)^{-1}
B_{\mbox{\rm {\tiny PT}}}(k)$ converges 
to $B_{0}(0)$ on $D(H_{0})$ as $k\to 0$.
\end{description}
If there exists a ground state 
$\Psi_{\mbox {\rm {\tiny QFT}}}$ 
such that  
\begin{eqnarray}
\frac{1}{\omega(\cdot)}
\left( \Phi\, ,\, 
( \widehat{H}_{\mbox{\rm {\tiny QFT}}}  
+ \omega(\cdot))^{-1}\widehat{H}_{\mbox{\rm {\tiny QFT}}}
B_{\mbox{\rm {\tiny PT}}}(\cdot)
\Psi_{\mbox{\rm {\tiny QFT}}}
\right)_{\mathcal F} 
\in  L^{2}(\mathbb{R}^{d})
\label{assump:zero}
\end{eqnarray}
for a vector $\Phi \in  
D(T_{\mbox{\rm {\tiny PT}}})$, 
then $\left( \Phi\, ,\, 
B_{0}(0)\Psi_{\mbox{\rm {\tiny QFT}}}\right)_{\mathcal F} 
= 0$. 
\end{theorem}

\demo 
Suppose there exists a ground state 
$\Psi_{\mbox{\rm {\tiny QFT}}}$ 
satisfying the condition (\ref{assump:zero}) 
for a vector  
$\Phi \in  D(T_{\mbox{\rm {\tiny PT}}})$. 
Since $\Phi \in  D(T_{\mbox{\rm {\tiny PT}}})$, 
we have 
$$ 
\left(\Phi\, ,\, 
(\widehat{H}_{\mbox{\rm {\tiny QFT}}} + 
\omega(\cdot))^{-1}B_{\mbox{\rm {\tiny PT}}}(\cdot)
\Psi_{\mbox{\rm {\tiny QFT}}}\right)_{\mathcal F} 
\in  L^{2}(\mathbb{R}^{d})
$$
by the definition 
(\ref{eq:Carleman-domain}). 
So, we can define $F \in  L^{2}({\mathbb R}^{d})$ 
by 
\begin{eqnarray*}
F(k) 
&:=&  
\left(\Phi\, ,\, 
(\widehat{H}_{\mbox{\rm {\tiny QFT}}} + 
\omega(k))^{-1}B_{\mbox{\rm {\tiny PT}}}(k)
\Psi_{\mbox{\rm {\tiny QFT}}}\right)_{\mathcal F} \\ 
&{}& \qquad 
+\, \frac{1}{\omega(k)}
\left( \Phi\, ,\, 
( \widehat{H}_{\mbox{\rm {\tiny QFT}}}  
+ \omega(k))^{-1}\widehat{H}_{\mbox{\rm {\tiny QFT}}}
B_{\mbox{\rm {\tiny PT}}}(k)
\Psi_{\mbox{\rm {\tiny QFT}}}
\right)_{\mathcal F},  
\end{eqnarray*}
where we used the condition (\ref{assump:zero}) 
in the second term of RHS. 
Since $(\widehat{H}_{\mbox{\tiny QFT}} + 
\omega(k))^{-1}B_{\mbox{\rm {\tiny PT}}}(k)$ is bounded 
for every $k \in \mathbb{R}^{d}\setminus \{ 0\}$ 
by (Ass.2), 
we have 
\begin{eqnarray}
\nonumber 
&{}& 
\left[B_{\mbox{\rm {\tiny PT}}}(k)\, ,\, 
(\widehat{H}_{\mbox{\tiny QFT}} + 
\omega(k))^{-1}\right] \\ 
&=& 
(\widehat{H}_{\mbox{\rm {\tiny QFT}}} + 
\omega(k))^{-1} 
\left[ 
\widehat{H}_{\mbox{\rm {\tiny QFT}}}\, ,\, 
B_{\mbox{\rm {\tiny PT}}}(k)
\right]
(\widehat{H}_{\mbox{\rm {\tiny QFT}}} + 
\omega(k))^{-1}.  
\label{eq:KEY}
\end{eqnarray}
So, Eq. (\ref{eq:KEY}) leads us to:   
$$F(\cdot) 
= 
\frac{1}{\omega(\cdot)}
\left(\Phi\, ,\, B_{\mbox{\rm {\tiny PT}}}(\cdot)
\Psi_{\mbox{\rm {\tiny QFT}}}\right)_{\mathcal F} 
$$
as an identity on $L^{2}({\mathbb R}^{d}_{>\varepsilon})$ 
for every $\varepsilon > 0$. 
Set $B_{0}(k)$ as $B_{0}(k) := 
\lambda(k)^{-1}B_{\mbox{\rm {\tiny PT}}}(k)$ 
for $k \in  \mathbb{R}^{d}$ with 
$0 < |k| < \varepsilon_{0}$. 
Then, the above equation implies  
$$
\int_{\varepsilon<|k|<\varepsilon_{0}}
\frac{|\lambda(k)|^{2}}{\omega(k)^{2}}
\biggl| 
\left(\Phi\, ,\, B_{0}(k)
\Psi_{\mbox{\rm {\tiny QFT}}}\right)_{\mathcal F}
\biggr|^{2}dk 
= 
\int_{\varepsilon<|k|<\varepsilon_{0}}|F(k)|^{2}dk. 
$$ 
Since $F \in  L^{2}({\mathbb R}^{d})$, 
taking $\varepsilon\to 0$ and using 
Lebesgue's monotone convergence theorem, we have 
\begin{eqnarray}
\frac{\lambda(\cdot)}{\omega(\cdot)}
\left(\Phi\, ,\, B_{0}(\cdot)
\Psi_{\mbox{\rm {\tiny QFT}}}\right)_{\mathcal F} 
\in  L^{2}({\mathbb R}_{<\varepsilon_{0}}^{d}). 
\label{eq:2/7-1}
\end{eqnarray}
Since $|\left( \Phi\, ,\, B_{0}(k)
\Psi_{\mbox{\rm {\tiny QFT}}}\right)_{\mathcal F}|^{2} 
\to |\left( \Phi\, ,\, B_{0}(0)
\Psi_{\mbox{\rm {\tiny QFT}}}\right)_{\mathcal F}|^{2}$ 
as $k \to 0$ by the assumption (3), 
for every $\varepsilon > 0$ there exists 
a positive number $\delta_{\Phi}(\varepsilon) > 0$ 
such that 
$$ 
|\left( \Phi\, ,\, B_{0}(0)
\Psi_{\mbox{\rm {\tiny QFT}}}\right)_{\mathcal F}|^{2} 
- \varepsilon 
\le 
|\left( \Phi\, ,\, B_{0}(k)
\Psi_{\mbox{\rm {\tiny QFT}}}\right)_{\mathcal F}|^{2}\quad  
$$
for every $k$ with 
$|k| < \delta_{\Phi}(\varepsilon)$. 
Set $\varepsilon_{0}\wedge \delta_{\Phi}(\varepsilon)$ 
as $\varepsilon_{0}\wedge \delta_{\Phi}(\varepsilon) 
:= \min\{\varepsilon_{0},\delta_{\Phi}(\varepsilon)\}$.  
Then, we have 
\begin{eqnarray*}
&{}& 
\left\{
\biggl| 
\left( \Phi\, ,\, B_{0}(0)
\Psi_{\mbox{\rm {\tiny QFT}}}\right)_{\mathcal F}
\biggr|^{2} - \varepsilon
\right\} 
\int_{|k|<\varepsilon_{0}\wedge \delta_{\Phi}(\varepsilon)}
\frac{|\lambda(k)|^{2}}{\omega(k)^{2}}dk \\ 
&\le& 
\int_{|k|<\varepsilon_{0}}
\frac{|\lambda(k)|^{2}}{\omega(k)^{2}}
\biggl| 
\left( \Phi\, ,\, B_{0}(k)
\Psi_{\mbox{\rm {\tiny QFT}}}\right)_{\mathcal F}
\biggr|^{2}dk 
< \infty 
\end{eqnarray*}
by the condition (\ref{eq:2/7-1}). 
So, by the assumption (2) we are bound to conclude that 
$|\left( \Phi\, ,\, B_{0}(0)
\Psi_{\mbox{\tiny QFT}}\right)_{\mathcal F}
|^{2}  \le  \varepsilon$. 
Thus, taking $\varepsilon\to 0$ yields   
$\left( \Phi\, ,\, B_{0}(0)
\Psi_{\mbox{\rm {\tiny QFT}}}\right)_{\mathcal F} = 0$.  
\qed

We state a useful domain property 
which causes the absence of the mass gap. 
Let $R(\widehat{H}_{\mbox{\rm {\tiny QFT}}})$ 
denote the range of $\widehat{H}_{\mbox{\rm {\tiny QFT}}}$: 
\begin{eqnarray*}
R(\widehat{H}_{\mbox{\rm {\tiny QFT}}}) 
:= 
\left\{
\widehat{H}_{\mbox{\rm {\tiny QFT}}}\Psi \, \Big|\, 
\Psi \in  D(\widehat{H}_{\mbox{\rm {\tiny QFT}}})
\right\}. 
\end{eqnarray*}

\begin{theorem}
\label{lemma:megane1}
Assume there is a ground state 
of $H_{\mbox{\rm {\tiny QFT}}}$. 
If $B_{\mbox{\rm {\tiny PT}}}(k)$ is a bounded 
operator on $\mathcal{F}$ for every $k \in  
\mathbb{R}^{d}\setminus\{ 0\}$ such that 
$\| B_{\mbox{\rm {\tiny PT}}}(\cdot)
\|_{\mathcal{B}(\mathcal{F})} \in  
L^{2}(\mathbb{R}^{d})$, then 
$R(\widehat{H}_{\mbox{\rm {\tiny QFT}}}) 
\subset 
D(T_{\mbox{\rm {\tiny PT}}})$. 
\end{theorem}

\demo 
For every $\Phi \in  R(\widehat{H}_{\mbox{\rm {\tiny QFT}}})$, 
there is a $\Psi \in  D(\widehat{H}_{\mbox{\rm {\tiny QFT}}})$ 
such that $\Phi = \widehat{H}_{\mbox{\rm {\tiny QFT}}}\Psi$. 
Thus, for every $\varepsilon > 0$ we have the following estimate: 
\begin{eqnarray*}
&{}& 
\int_{\varepsilon < |k| < \varepsilon^{-1}}
|\left( K_{\mbox{\rm {\tiny PT}}}(k)\, ,\, 
\Phi\right)_{\mathcal{F}}|^{2}dk \\ 
&=& 
\int_{\varepsilon < |k| < \varepsilon^{-1}}
\Bigl|\left( \widehat{H}_{\mbox{\rm {\tiny QFT}}}
(\widehat{H}_{\mbox{\rm {\tiny QFT}}}+\omega(k))^{-1}
B_{\mbox{\rm {\tiny PT}}}(k)
\Psi_{\mbox{\rm {\tiny QFT}}}\, ,\, 
\Psi\right)_{\mathcal{F}}
\Bigr|^{2}dk \\ 
&\le& 
\|\Psi\|_{\mathcal{F}}^{2}
\int_{\mathbb{R}^{d}}
\| B_{\mbox{\rm {\tiny PT}}}(k)
\|_{\mathcal{B}(\mathcal{F})}^{2}dk 
< \infty. 
\end{eqnarray*}
Therefore, taking $\varepsilon \to 0$, 
together with Lebesgue's monotone convergence theorem, 
yields $\Phi \in  D(T_{\mbox{\rm {\tiny PT}}})$. 
\qed

\section{IR Catastrophe and Absence of Grand State}
\label{sec:IRCAGS}

In the case where IR singularity of 
$T_{\mbox{\rm {\tiny PT}}}$ is determined 
by the singularity of the function $\lambda$ on $\mathbb{R}^{d}$ 
as seen in Theorem \ref{lemma:zero},  
we introduce a notion for the order of IR singularity 
of the maximal Carleman operator 
$T_{\mbox{\rm {\tiny PT}}}$ 
at $k = 0$. 
That is, in this section we assume that 
there is a measurable function $\lambda$ 
on $\mathbb{R}^{d}$ satisfying the condition (1) of 
Theorem \ref{lemma:zero} and an operator $B_{0}(k)$ 
acting in ${\mathcal F}$ for every $k \in  
{\mathbb R}^{d}$ such that 
the following (S1) and (S2) are satisfied:  
\begin{description}
\item[(S1)] 
$B_{\mbox{\rm {\tiny PT}}}(k) = 
\lambda(k)B_{0}(k)$ 
on $D(H_{0})$ 
for every $k \in  
{\mathbb R}^{d}\setminus \{ 0\}$;  
\item[(S2)] $B_{0}(k)\Psi \longrightarrow 
B_{0}(0)\Psi$ in $\mathcal{F}$
as $k \to 0$ for every $\Psi \in  D(H_{0})$.  
\end{description}

IR singularity condition \cite{ah2,ahh} is 
reinterpreted as $\lambda/\omega \notin  
L^{2}(\mathbb{R}^{d})$ for the Carleman operator 
$T_{\mbox{\rm {\tiny PT}}}$. 
We extend the notion of IR singularity condition. 
Our new notion is: 

\begin{definition}
\label{def:IRSC}
{\rm We say $\omega$ and $\lambda$ satisfy 
IR singularity condition  
if there are constants $\gamma_{1}^{}, \gamma_{2}^{}, 
\varepsilon_{2}^{} > 0$ 
such that $\lambda/\omega^{\gamma} 
\in  L^{2}(\mathbb{R}^{d})$ 
for every $\gamma$ with $\gamma < \gamma_{1}^{}$ 
and $\lambda/\omega^{\gamma} 
\notin  L^{2}(\mathbb{R}^{d}_{<\varepsilon})$ 
for every $\gamma$ and $\varepsilon$ 
with $\gamma > \gamma_{2}^{}$ and $\varepsilon_{2}^{} 
\ge \varepsilon > 0$.  
We say $\gamma$ is in the IR-safe region 
(resp. the IR-divergent region) if $\gamma < \gamma_{1}^{}$ 
(resp. $\gamma > \gamma_{2}^{}$). 
In particular, we call $\gamma_{\mathrm{c}}^{}$ 
the order of IR singularity condition when 
$\gamma_{1}^{} = \gamma_{2}^{} = \gamma_{\mathrm{c}}^{}$ 
and $\lambda/\omega^{\gamma_{\mathrm{c}}^{}} \notin  
L^{2}(\mathbb{R}^{d}_{<\varepsilon})$ 
for every $\varepsilon$ 
with $\varepsilon_{2}^{} 
\ge \varepsilon > 0$.  
In this case, we also say $\gamma = 
\gamma_{\mathrm{c}}^{}$ is in the IR-divergent region.
} 
\end{definition}
 
\begin{example}
\label{example:IRSC-standard}
Most standard assumptions for $\omega$ and $\lambda$ 
are $\lambda/\sqrt{\omega} \in  L^{2}(\mathbb{R}^{d})$ 
and $\lambda/\omega \notin  L^{2}(\mathbb{R}^{d})$. 
So, in this case, $\omega$ and $\lambda$ satisfy 
IR singularity condition and $1/2 < \gamma_{\mathrm{c}}^{} 
\le 1$.   
\end{example}

We say {\it a symmetric operator $S$ 
strongly commutes with $H_{\mbox{\rm {\tiny QFT}}}$} 
if $e^{itH_{\mbox{\rm {\tiny QFT}}}}S \subset 
Se^{itH_{\mbox{\rm {\tiny QFT}}}}$ 
for all $t \in  \mathbb{R}$. 
Then, we can derive the following theorem from 
Theorem \ref{theorem:hougan}. 
This is a generalization of 
Derezi\'{n}ski and G\'{e}rard's 
\cite[Lemma 2.6]{dg} and ours \cite[Theorem 3.4]{ahh}.  

\begin{theorem} 
\label{theorem:absence-base}
Suppose $D(H_{\mbox{\rm {\tiny QFT}}}) 
= D(H_{0})$.  
Assume $\omega$ and $\lambda$ satisfy 
IR singularity condition with 
the order $\gamma_{\mathrm{c}}^{}$ less 
than or equal to $1$ (i.e., 
$\gamma_{\mathrm{c}}^{} \le 1$).  
Then, there is no ground state 
$\Psi_{\mbox{\rm {\tiny QFT}}}$ 
satisfying all of the following (i)--(iii): 
\begin{description}
\item[(i)] 
$B_{0}(0)$ is symmetric and 
strongly commutes with $H_{\mbox{\rm {\tiny QFT}}}$. 
\item[(ii)] $B_{0}(0)
\Psi_{\mbox{\rm {\tiny QFT}}} \ne 0$. 
\item[(iii)] $\sup_{k\in\mathbb{R}^{d}}
\omega(k)^{\gamma-1}
\|(B_{0}(k) - B_{0}(0))\Psi_{\mbox{\rm {\tiny QFT}}}
\|_{\mathcal{F}} < \infty$ for some 
$\gamma$ in the IR-safe region. 
\end{description}

\end{theorem}

\demo 
We use the reduction of absurdity. 
So, we suppose that a ground state 
$\Psi_{\mbox{\tiny QFT}}$ exists such that 
all of (i)--(iii) hold. 
For all $\Phi \in  D(N^{1/2})$ and 
every $k \in \mathbb{R}^{d}\setminus \{0\}$ 
we have 
\begin{eqnarray*}
\left( K_{\mbox{\rm {\tiny PT}}}(k)\, ,\, 
\Phi\right)_{\mathcal F} 
&=& 
\lambda(k) 
\left( 
(\widehat{H}_{\mbox{\rm {\tiny QFT}}} 
+ \omega(k))^{-1}B_{0}(0)\Psi_{\mbox{\rm {\tiny QFT}}}\, 
,\, \Phi\right)_{\mathcal F} \\ 
&{}&  
+ 
\lambda(k) 
\left( 
(\widehat{H}_{\mbox{\rm {\tiny QFT}}} 
+ \omega(k))^{-1}(B_{0}(k) - B_{0}(0))\Psi_{\mbox{\rm {\tiny QFT}}}\, 
,\, \Phi\right)_{\mathcal F} \\ 
&=& 
\frac{\lambda(k)}{\omega(k)} 
\left( 
B_{0}(0)\Psi_{\mbox{\rm {\tiny QFT}}}\, 
,\, \Phi\right)_{\mathcal F} \\ 
&{}&  
+ 
\lambda(k) 
\left( 
(\widehat{H}_{\mbox{\rm {\tiny QFT}}} 
+ \omega(k))^{-1}(B_{0}(k) - B_{0}(0))\Psi_{\mbox{\rm {\tiny QFT}}}\, 
,\, \Phi\right)_{\mathcal F} 
\end{eqnarray*}
by (i). 
This equation implies 
\begin{eqnarray*}
0 &\le& 
|\left( B_{0}(0)\Psi_{\mbox{\tiny QFT}}\, 
,\, \Phi\right)_{\mathcal{F}}|^{2}
\int_{|k|> \varepsilon}
\frac{|\lambda(k)|^{2}}{\omega(k)^{2}}dk \\ 
&\le&  
2 \int_{{\mathbb R}^{d}} 
|\left( K_{\mbox{\tiny PT}}(k)\, ,\, \Phi
\right)_{\mathcal F}|^{2}dk \\ 
&{}&
+ 
2\|\Phi\|_{\mathcal{F}}^{2} 
\left(
\sup_{k\in\mathbb{R}^{d}}
\omega(k)^{\gamma-1}
\|(B_{0}(k) - B_{0}(0))\Psi_{\mbox{\rm {\tiny QFT}}}
\|_{\mathcal{F}}
\right)^{2}
\int_{{\mathbb R}^{d}} 
\frac{|\lambda(k)|^{2}}{\omega(k)^{2\gamma}}
dk. 
\end{eqnarray*}
Here we note that the two integrals 
of RHS are finite 
by Theorem \ref{theorem:hougan} and (iii), and 
they are independent of $\varepsilon > 0$. 
Taking $\varepsilon \to 0$, 
Lebesgue's monotone convergence theorem 
tells us that $\left( B_{0}(0)\Psi_{\mbox{\tiny QFT}}\, 
,\, \Phi\right)$ 
is bound to be $0$ 
(i.e., $\left( B_{0}(0)\Psi_{\mbox{\tiny QFT}}\, 
,\, \Phi\right) = 0$) for all $\Phi \in  D(N^{1/2})$ 
since $\lambda/\omega \notin L^{2}(\mathbb{R}^{d})$. 
Since $D(N^{1/2})$ is dense in $\mathcal{F}$, 
we reach $B_{0}(0)\Psi_{\mbox{\tiny QFT}} = 0$ finally, 
which contradicts (ii). 
\qed

We can obtain 
Derezi\'{n}ski and G\'{e}rard's 
\cite[Lemma 2.6]{dg} 
as a corollary of 
Theorem \ref{theorem:absence-base}: 

\begin{corollary}
\label{proposition:absence1-1}
Suppose $D(H_{\mbox{\rm {\tiny QFT}}}) 
= D(H_{0})$ and $B_{\mbox{\rm {\tiny PT}}}(k)$ can be 
decomposed into $B_{\mbox{\rm {\tiny PT}}}(k) 
= g(k)I\otimes I + J_{\mathrm {err}}(k)$ 
on $D(H_{0})$ for every $k \in  \mathbb{R}^{d}\setminus\{ 0\}$. 
Assume the following (1)--(3): 
\begin{description}
\item[(1)] $g/\omega \notin  L^{2}(\mathbb{R}^{d})$,  
\item[(2)] $g/\omega^{\gamma_{0}^{}} \in  L^{2}(\mathbb{R}^{d})$ 
for a $\gamma_{0}^{}$ with 
$0 < \gamma_{0}^{} < 1$, 
\item[(3)] $g(k)^{-1}J_{\mathrm{err}}(k)\Psi 
\longrightarrow 0$ as $k \to 0$ for every $\Psi \in  D(H_{0})$. 
\end{description}
Then, there is no ground state 
$\Psi_{\mbox{\rm {\tiny QFT}}}$ satisfying 
\begin{eqnarray}
\sup_{k\in\mathbb{R}^{d}}
\omega(k)^{\gamma_{0}^{}-1}g(k)^{-1}
\|J_{\mathrm {err}}(k)
\Psi_{\mbox{\rm {\tiny QFT}}}
\|_{\mathcal{F}} < \infty. 
\label{eq:DG}
\end{eqnarray}
\end{corollary}

\demo 
Set $B_{0}(0)$, $\lambda(k)$, and $B_{0}(k)$ as
$B_{0}(0) := I\otimes I$, 
$\lambda(k) := g(k)$, and 
$B_{0}(k) := 
\lambda(k)^{-1}J_{\mathrm{err}}(k) + I\otimes I$, 
respectively. Then, the assumption (3) implies 
(S1), (S2), and (i) and (ii) of Theorem 
\ref{theorem:absence-base}. 
The assumptions (1) and (2) 
tell us that $\omega$ and $\lambda$ satisfy IR 
singularity condition such that $1$ is in the 
IR-divergent region and $\gamma_{0}^{}$ in the 
IR-safe region. 
Thus, Theorem \ref{theorem:absence-base} concludes 
that there is no ground sate satisfying 
\begin{eqnarray*} 
&{}& 
\sup_{k\in\mathbb{R}^{d}}
\omega(k)^{\gamma_{0}^{}-1}g(k)^{-1}
\|J_{\mathrm {err}}(k)
\Psi_{\mbox{\rm {\tiny QFT}}}
\|_{\mathcal{F}} \\ 
&=&  
\sup_{k\in\mathbb{R}^{d}}\omega(k)^{\gamma_{0}^{}-1}
\|(B_{0}(k) - B_{0})\Psi_{\mbox{\rm {\tiny QFT}}}
\|_{\mathcal{F}} 
< \infty. 
\end{eqnarray*} 
\qed

As a corollary of Theorem \ref{theorem:absence-base} 
we also obtain \cite[Theorem 3.4]{ahh} of which statement 
can be applied to several models as well as in \cite{ahh}:  

\begin{corollary}
\label{proposition:absence0}
Suppose $D(H_{\mbox{\rm {\tiny QFT}}}) 
= D(H_{0})$ and there is an operator 
$C_{\mbox{\rm {\tiny PT}}}$ with 
$D(C_{\mbox{\rm {\tiny PT}}}) 
\supset D(H_{0})$ such that 
$B_{\mbox{\rm {\tiny PT}}}(k) 
= \lambda(k)C_{\mbox{\rm {\tiny PT}}}$ 
on $D(H_{0})$ for 
all $k \in  \mathbb{R}^{d}$. 
Assume the following (1)--(3): 
\begin{description}
\item[(1)] $\lambda/\omega \notin L^{2}(\mathbb{R}^{d})$, 
\item[(2)] $\lambda/\omega^{\gamma_{0}^{}} \in  L^{2}(\mathbb{R}^{d})$ 
for a $\gamma_{0}^{}$ with $0 < \gamma_{0}^{} < 1$,
\item[(3)] $C_{\mbox{\rm {\tiny PT}}}$ 
is symmetric and strongly commutes with 
$H_{\mbox{\rm {\tiny QFT}}}$. 
\end{description}
Then, there is no ground state $\Psi_{\mbox{\rm {\tiny QFT}}}$ 
satisfying $C_{\mbox{\rm {\tiny PT}}}\Psi_{\mbox{\rm {\tiny QFT}}} 
\ne 0$. 
\end{corollary}

\demo 
Set $B_{0}(k)$ as 
$B_{0}(k) := C_{\mbox{\rm {\tiny PT}}}$ 
for all $k \in \mathbb{R}^{d}$. 
Then, (S1), (S2), and (i) of Theorem 
\ref{theorem:absence-base} hold 
by the assumption (3). 
The assumption (1) implies $\gamma_{\mathrm{c}}^{} 
\le 1$.  
Since $B_{0}(k) - B_{0}(0) = 0$ on $D(H_{0})$ now, 
the condition (iii) in Theorem \ref{theorem:absence-base} 
always holds for $\gamma_{0}^{}$ in the IR-safe region 
by the assumption (2). 
Thus, Theorem \ref{theorem:absence-base} 
leads us to the conclusion that 
there is no ground state $\Psi_{\mbox{\rm {\tiny QFT}}}$ 
satisfying $C_{\mbox{\rm {\tiny PT}}}\Psi_{\mbox{\tiny QFT}} 
= B_{0}(0)\Psi_{\mbox{\tiny QFT}} 
\ne 0$.
\qed

The following theorem follows 
from Theorem \ref{lemma:zero}:

\begin{theorem}
\label{corollary:absence2-2}
Assume $D(H_{\mbox{\rm {\tiny QFT}}}) = D(H_{0})$ 
and $\lambda/\omega \notin  
L^{2}(K)$ for every neighborhood $K$ 
of $k=0$. 
Then, there is no ground state 
$\Psi_{\mbox{\rm {\tiny QFT}}}$ in 
$D(T_{\mbox{\rm {\tiny PT}}})$ satisfying 
$\langle B_{0}(0)\rangle_{\mathrm{gs}} 
\ne 0$. 
Thus, in particular, if $B_{0}(0)\Psi \ne 0$ 
for every $\Psi \in  D(H_{0})$, 
then no ground state exists in 
$D(T_{\mbox{\rm {\tiny PT}}})$, namely, 
IR catastrophe occurs. 
\end{theorem}

\demo 
Let us suppose there is a ground state 
$\Psi_{\mbox{\rm {\tiny QFT}}}$ in 
$D(T_{\mbox{\rm {\tiny PT}}})$ now. 
We easily have  
$$
\frac{1}{\omega(\cdot)}
\left( \Psi_{\mbox{\rm {\tiny QFT}}}\, ,\, 
( \widehat{H}_{\mbox{\tiny QFT}}  
+ \omega(\cdot))^{-1}
\widehat{H}_{\mbox{\rm {\tiny QFT}}}
B_{\mbox{\tiny PT}}(\cdot)
\Psi_{\mbox{\rm {\tiny QFT}}}
\right)_{\mathcal F} = 0.  
$$ 
Thus, it follows immediately 
from Theorem \ref{lemma:zero} 
that $\langle B_{0}(0)\rangle_{\mathrm{gs}}$ 
$=$ $\left( \Psi_{\mbox{\rm {\tiny QFT}}}\, ,\,\right.$  
$B_{0}(0)\Psi_{\mbox{\rm {\tiny QFT}}}$
$\left.\right)_{\mathcal F}$ $=$ $0$, 
which means our theorem holds.
\qed 

\begin{remark}
\label{rem:PF}
As mentioned in Section \ref{section:Intro} 
(i.e., \cite[p.212 and p.213]{hirokawa-e}), 
since the full model of non-relativistic quantum 
electrodynamics \cite{bfs2,gll} has local gauge invariance, 
the commutation relation 
$i[ H_{\mbox{\rm {\tiny QFT}}}\, ,\, x] = v$ 
holds for the full model, 
where $x$ and $v$ are the position 
and velocity of the non-relativistic electron. 
This commutation relation cancels IR singularity 
in Eqs.(\ref{eq:establishment}) and (\ref{eq:PT}). 
Thus, IR catastrophe does not occur for the full model. 
Since $B_{0}(0) = v$ and the above commutation relation 
implies $\langle v\rangle_{\mathrm{gs}} = 0$ 
for the full model, we have 
$\langle B_{0}(0)\rangle_{\mathrm{gs}} = 0$. 
Namely, local gauge invariance also works 
to avoid IR catastrophe in Theorem \ref{corollary:absence2-2}. 
\end{remark}

\begin{remark}
\label{rem:IRD}
Theorem \ref{corollary:absence2-2} is a general expression 
of IR catastrophe for the Nelson Hamiltonian 
\cite{betz,hirokawa-nelson} and for the spin-boson Hamiltonian 
\cite{ahh,ahh2,hh}. 
\end{remark}

Let us denote by $\sigma_{\mathrm{ess}}(S)$ 
the essential spectrum of a self-adjoint 
operator $S$. 

Combining Lemma \ref{lemma:megane1} and 
Theorem \ref{corollary:absence2-2} 
yields the following theorem, which states 
IR singularity condition prohibits 
$H_{\mbox{\rm {\tiny QFT}}}$ 
from making the mass gap: 

\begin{theorem}
\label{theorem:megane2}
Suppose $D(H_{\mbox{\rm {\tiny QFT}}}) = D(H_{0})$. 
Assume the following (1)--(3):  
\begin{description}
\item[(1)] $\lambda/\omega \notin  
L^{2}(\mathbb{R}^{d})$, 
\item[(2)] $B_{\mbox{\rm {\tiny PT}}}(k)$ is 
a bonded operator acting on $\mathcal{F}$ for 
every $k \in  \mathbb{R}^{d}\setminus\{ 0\}$ 
such that $\| B_{\mbox{\rm {\tiny PT}}}(\cdot)
\|_{\mathcal{B}(\mathcal{F})} \in  
L^{2}(\mathbb{R}^{d})$, 
\item[(3)] $B_{0}(0)\Psi \ne 0$ 
for every $\Psi \in  D(H_{0})$.  
\end{description}
Then, there is no gap between the ground state 
energy and the infimum of the essential spectrum 
of $H_{\mbox{\rm {\tiny QFT}}}$: 
$$
E_{0}(H_{\mbox{\rm {\tiny QFT}}}) 
= \inf\sigma_{\mathrm{ess}}(H_{\mbox{\rm {\tiny QFT}}}).
$$ 
\end{theorem}

\demo 
We prove our statement by the reduction of absurdity. 
So, suppose $E_{0}(H_{\mbox{\rm {\tiny QFT}}})$ 
$<$ $\inf\sigma_{\mathrm{ess}}(H_{\mbox{\rm {\tiny QFT}}})$.
Then, there is a ground state $\Psi_{\mbox{\rm {\tiny QFT}}}$ 
and $0 < \inf\sigma_{\mathrm{ess}}
(\widehat{H}_{\mbox{\rm {\tiny QFT}}})$.
Thus, for every $\Phi \in  D(\widehat{H}_{\mbox{\rm {\tiny QFT}}}) 
\cap \mathrm{ker}(\widehat{H}_{\mbox{\rm {\tiny QFT}}})^{\perp}$, 
we have 
$\inf\sigma_{\mathrm{ess}}
(\widehat{H}_{\mbox{\rm {\tiny QFT}}})
\|\Phi\|_{\mathcal{F}} \le 
\|\widehat{H}_{\mbox{\rm {\tiny QFT}}}\Phi
\|_{\mathcal{F}}$. 
This inequality is equivalent to the fact 
that $R(\widehat{H}_{\mbox{\rm {\tiny QFT}}})$ 
is closed as well known. 
So, we have $\mathcal{F} = 
\mathrm{ker}(\widehat{H}_{\mbox{\rm {\tiny QFT}}}) 
\bigoplus R(\widehat{H}_{\mbox{\rm {\tiny QFT}}})$. 
Let $P_{0}$ be the orthogonal projection 
onto $\mathrm{ker}(\widehat{H}_{\mbox{\rm {\tiny QFT}}})$. 
For every $\Psi \in  \mathcal{F}$ 
with $P_{0}\Psi \ne 0$, there are $\Psi_{n} \in  
D(T_{\mbox{\rm {\tiny PT}}})$, $n \in  \mathbb{N}$, 
such that $\Psi_{n} \to  \Psi$ as $n\to\infty$, 
since $D(T_{\mbox{\rm {\tiny PT}}})$ is dense in 
$\mathcal{F}$ by Theorem \ref{theorem:hougan}. 
We have $P_{0}\Psi_{n} \ne 0$ for 
almost all $n \in  \mathbb{N}$ except finite $n$'s 
because $P_{0}\Psi \ne 0$. 
We note $\Psi_{n} = P_{0}\Psi_{n} 
+ (I - P_{0})\Psi_{n}$ and 
$(I - P_{0})\Psi_{n} \in  
R(\widehat{H}_{\mbox{\rm {\tiny QFT}}}) 
\subset D(T_{\mbox{\rm {\tiny PT}}})$ 
by Lemma \ref{lemma:megane1}. 
Thus, we obtain 
$0 \ne P_{0}\Psi_{n} = 
\Psi_{n} - (I - P_{0})\Psi_{n} 
\in  D(T_{\mbox{\rm {\tiny PT}}})$. 
On the other hand, 
$P_{0}\Psi_{n} \notin  D(T_{\mbox{\rm {\tiny PT}}})$ 
by Theorem \ref{corollary:absence2-2} 
since $P_{0}\Psi_{n} \ne 0$ is also a ground state 
of $\widehat{H}_{\mbox{\rm {\tiny QFT}}}$. 
Therefore, we reach a contradiction. 

\qed 

As explained in Example \ref{example:dame} below, 
we need another statement 
to avoid the restriction coming from (ii) 
in Theorem \ref{theorem:absence-base}. 
We take account of the order of IR singularity condition. 
Then, we obtain the following from Theorem \ref{theorem:hougan}: 

\begin{theorem}
\label{corollary:absence2-3}
Suppose $D(H_{\mbox{\rm {\tiny QFT}}}) = D(H_{0})$ 
and $B_{0}(0)$ is symmetric and strongly commutes with 
$H_{\mbox{\rm {\tiny QFT}}}$. 
Assume there is an $\varepsilon_{0} > 0$ and 
operators $B_{j}(k)$, $j=1, \cdots, d$, 
acting in $\mathcal{F}$ for every $k \in  
\mathbb{R}^{d}\setminus\{0\}$ such that 
$B_{0}(k)\Psi_{\mbox{\rm {\tiny QFT}}}$ 
is decomposed into 
$B_{0}(k)\Psi_{\mbox{\rm {\tiny QFT}}}$ 
$=$ $B_{0}(0)\Psi_{\mbox{\rm {\tiny QFT}}} 
+ \sum_{j=1}^{d}k_{j}
B_{j}(k)\Psi_{\mbox{\rm {\tiny QFT}}}$ 
for $|k| < \varepsilon_{0}^{}$. 
If $\omega$ and $\lambda$ satisfy IR 
singularity condition with the order $\gamma_{\mathrm{c}}^{}$, 
and moreover, 
$$
\int_{|k|<\varepsilon_{0}} 
\frac{|k_{j}||\lambda(k)|^{2}}{\omega(k)^{1+\gamma}}
dk < \infty 
$$ 
for a $\gamma > 0$ with $\gamma < \gamma_{\mathrm{c}}^{} 
< (1+\gamma)/2$ and $j = 1, \cdots, d$,
then there is no ground state 
$\Psi_{\mbox{\rm {\tiny QFT}}}$ satisfying 
$B_{0}(0)\Psi_{\mbox{\rm {\tiny QFT}}} \ne 0$ 
and $\sup_{|k|<\varepsilon_{0}}
\|B_{j}(k)\Psi_{\mbox{\rm {\tiny QFT}}}
\|_{\mathcal{F}} < \infty$ for 
all $j=1, \cdots, d$. 
\end{theorem}

\demo 
We use the reduction of absurdity. 
So, we suppose that there is such a 
ground state $\Psi_{\mbox{\rm {\tiny QFT}}}$. 
Let us fix $\Phi \in  D(N^{1/2})$ arbitrarily 
and define a function $F_{\Phi}(k)$ by 
$F_{\Phi}(k) := (K_{\mbox{\rm {\tiny PT}}}(k)\, ,\, 
\Phi)_{\mathcal{F}}$. 
We define another function 
$F_{\gamma}(k)$ by 
$F_{\gamma}(k) := \lambda(k)\omega(k)^{-\,\gamma}$.  
Then, we have $F_{\Phi} \in  L^{2}(\mathbb{R}^{d})$ 
by Theorem \ref{theorem:hougan} and 
$F_{\gamma} \in  L^{2}(\mathbb{R}^{d})$ 
by our assumption. 
For every $\varepsilon$ with 
$\varepsilon < 
\min\{\varepsilon_{0}^{},
\varepsilon_{2}^{}\} =: \varepsilon_{0}^{}\wedge
\varepsilon_{2}^{}$, 
where $\varepsilon_{2}^{}$ is in 
Definition \ref{def:IRSC}, we have 
\begin{eqnarray}
\nonumber 
&{}& 
\int_{\varepsilon<|k|<\varepsilon_{0}^{}\wedge
\varepsilon_{2}^{}} 
F_{\Phi}(k)F_{\gamma}(k)dk  \\ 
\nonumber 
&=&
\left( 
B_{0}(0)\Psi_{\mbox{\rm {\tiny QFT}}}\, ,\, 
\Phi\right)_{\mathcal{F}} 
\int_{\varepsilon<|k|<
\varepsilon_{0}^{}\wedge
\varepsilon_{2}^{}} 
\frac{|\lambda(k)|^{2}}{\omega(k)^{1+\gamma}}dk \\ 
\nonumber 
&{}& 
+ 
\sum_{j=1}^{d}
\int_{\varepsilon<|k|<
\varepsilon_{0}^{}\wedge
\varepsilon_{2}^{}} 
\frac{k_{j}|\lambda(k)|^{2}}{\omega(k)^{\gamma}} \\ 
&{}& \qquad\qquad 
\left( 
(\widehat{H}_{\mbox{\rm {\tiny QFT}}} 
+ \omega(k))^{-1}B_{j}(k)\Psi_{\mbox{\rm {\tiny QFT}}}\, ,\, 
\Phi\right)_{\mathcal{F}}dk.  
\label{eq:new-method-1}
\end{eqnarray}
In the first term of RHS of the above, 
we used the assumption that $B_{0}(0)$ commutes 
with $H_{\mbox{\rm {\tiny QFT}}}$. 
We can estimate the last integrals as: 
\begin{eqnarray}
\nonumber 
&{}&
\Biggl|
\int_{\varepsilon<|k|<
\varepsilon_{0}^{}\wedge
\varepsilon_{2}^{}} 
\frac{k_{j}|\lambda(k)|^{2}}{\omega(k)^{\gamma}}
\left( 
(\widehat{H}_{\mbox{\rm {\tiny QFT}}} 
+ \omega(k))^{-1}B_{j}(k)\Psi_{\mbox{\rm {\tiny QFT}}}\, ,\, 
\Phi\right)_{\mathcal{F}}dk
\Biggr| \\ 
&\le& 
\|\Phi\|_{\mathcal{F}}
\sup_{|k|<\varepsilon_{0}^{}}
\|B_{j}(k)\Psi_{\mbox{\rm {\tiny QFT}}}
\|_{\mathcal{F}} 
\int_{|k|<\varepsilon_{0}^{}} 
\frac{|k_{j}||\lambda(k)|^{2}}{\omega(k)^{1+\gamma}}
dk < \infty.  
\label{eq:new-method-2}
\end{eqnarray}
Combining Eq.(\ref{eq:new-method-1}) 
and the inequality (\ref{eq:new-method-2}) 
gives us the inequality: 
\begin{eqnarray*}
0&\le& 
\Bigl|
\left( 
B_{0}(0)\Psi_{\mbox{\rm {\tiny QFT}}}\, ,\, 
\Phi\right)_{\mathcal{F}}\Bigr| 
\int_{\varepsilon<|k|<\min\{\varepsilon_{0}^{},
\varepsilon_{2}^{}\}} 
\frac{|\lambda(k)|^{2}}{\omega(k)^{1+\gamma}}dk \\ 
&\le& 
\| F_{\Phi}\|_{L^{2}(\mathbb{R}^{d})}
\| F_{\gamma}\|_{L^{2}(\mathbb{R}^{d})} \\ 
&{}& 
+ 
\|\Phi\|_{\mathcal{F}}
\sum_{j=1}^{d}\sup_{|k|<\varepsilon_{0}^{}}
\|B_{j}(k)\Psi_{\mbox{\rm {\tiny QFT}}}
\|_{\mathcal{F}} 
\int_{|k|<\varepsilon_{0}^{}} 
\frac{|k_{j}||\lambda(k)|^{2}}{\omega(k)^{1+\gamma}}
dk < \infty.  
\end{eqnarray*}
Taking $\varepsilon \to 0$, 
Lebesgue's monotone convergence theorem 
tells us that $\left( B_{0}(0)\Psi_{\mbox{\tiny QFT}}\, 
,\, \Phi\right)_{\mathcal{F}}$ 
is bound to be $0$ 
(i.e., $\left( B_{0}(0)\Psi_{\mbox{\tiny QFT}}\, 
,\, \Phi\right)_{\mathcal{F}} = 0$) 
for all $\Phi \in  D(N^{1/2})$ 
since $\lambda/\omega^{(1+\gamma)/2} \notin L^{2}(\mathbb{R}^{d})$. 
Since $D(N^{1/2})$ is dense in $\mathcal{F}$, 
we reach $B_{0}(0)\Psi_{\mbox{\tiny QFT}} = 0$ finally. 
This is a contradiction. 
\qed

\section{An Application} 
\label{sec:examples}

In this section, we consider the model of a non-relativistic 
electron coupled with a Bose field made from 
several sorts of phonons 
\cite[Chap.4]{kittel} 
or polaritons \cite[\S 11.4]{IL} in a material 
such as a crystal or a metal. 
Then, the order of IR singularity condition depends 
on the sorts of phonons or polaritons. 
Because each dispersion relation $\omega(k)$ 
is determined by an individual dispersion equation 
derived from the equation of motion of 
atoms in the material 
(see \cite{IL,kittel} 
for theoretical understanding 
and \cite{aho,ffk,HWSOPI,sjfddp,sa,yty} 
for experimental understanding). 
In addition, of course, the interaction function 
$\rho(k)$ depends on the property of the material. 
As in Eq.(\ref{eq:concrete-case}) of Example \ref{example:dame}, 
we idealize $\omega(k)$ and $\rho(k)$ 
in Theorem \ref{theorem:criterion} 
mathematically to investigate the order of IR singularity.  

We put the non-relativistic electron in the material. 
We suppose that the electron is negatively charged 
and thus is attracted by a plus-charged source 
which is caused by the positively charged ion cores 
caused by, for instance, the crystal lattice 
deformation \cite[\S 10.3]{IL} 
(also called the crystal lattice 
distortion \cite{eetal,feynman}). 
Thus, as the operator $A$ in Eq.(\ref{eq:Free}) 
we employ a Hamiltonian $H_{\mathrm{at}}$ given by 
the Schr\"{o}dinger operator 
with a potential $V$:  
\begin{eqnarray*}
H_{\mathrm {at}} \equiv   
\frac{1}{2}p^2  + V
\label{eq:A-GLP}
\end{eqnarray*} 
acting in ${\mathcal H} = L^{2}({\mathbb R}^{d})$, 
where $p := -i\nabla_{x}$ is the momentum of the electron. 
We use the natural units here.

As in \cite{hirokawa-nelson} 
we consider potentials $V$ in the class 
either (N1-1) or (N1-2) below. 
Here we say that \textit{$V$ is in class (N1-1)} 
(resp. {\it (N1-2)}) 
if the following (N1-1-1) and (N1-1-2) 
(resp. (N1-2-1) and (N1-2-2)) hold. 
These conditions are set so that if $V$ is in class 
(N1-1) or (N1-2), then $H_{\mathrm {at}}$ 
becomes a self-adjoint operator 
bounded from below with $D(H_{\mathrm {at}}) 
\subset D(p^{2})$, and moreover, 
$H_{\mathrm {at}}$ has a ground state $\psi_{\mathrm {at}}$. 
When we say that we assume (N1), 
we mean that either (N1-1) or (N1-2) is assumed.

\begin{description}
\item[(N1-1)]\cite{arai}: 
\item[(N1-1-1)] $H_{\mathrm {at}}$ is self-adjoint 
on $D(H_{\mathrm {at}}) 
\equiv D(p^{2})\cap D(V)$ and bounded from below, 
\item[(N1-1-2)] there exist positive constants 
$c_{1}$ and $c_{2}$ such that 
$|x|^{2} 
\le c_{1}V(x) + c_{2}$ 
for almost every (a.e.) $x \in  {\mathbb R}^{d}$, and 
${\displaystyle \int_{|x| \le R}|V(x)|^{2}dx < \infty}$  
for all $R > 0$.  
\end{description}

\begin{description}
\item[(N1-2)]  \cite{spohn}:  
\item[(N1-2-1)] $V \in  L^{2}({\mathbb R}^{d}) + 
L^{\infty}({\mathbb R}^{d})$ and $\lim_{|x|\to\infty}
|V(x)| = 0$. 
\end{description}
Following \cite[Theorem X15]{rs2} 
and \cite[\S XIII.4, Example 6]{rs4}  
the condition (N1-2-1) implies 
that $H_{\mathrm {at}}$ is self-adjoint on $D(p^{2})$; 
$V$ is infinitesimally $p^{2}$-bounded; and 
the essential spectrum 
$\sigma_{\mathrm{ess}}(H_{\mathrm {at}})$ 
of $H_{\mathrm {at}}$ is equal to 
$\left[\left. 0\, ,\, \infty\right)\right.$.  
So, we assume the following in addition: 

\begin{description}
\item[(N1-2-2)] $H_{\mathrm {at}}$ has a ground state 
$\psi_{\mathrm {at}}$ satisfying 
${\psi_{\mathrm {at}}} (x) > 0$ for  
a.e. $x \in {\mathbb R}^{d}$ and  
$E_{\mathrm {at}} := 
inf\sigma(H_{\mathrm {at}}) < 0$.  
\end{description}

In order to define the interaction Hamiltonian 
$H_{{\mathrm I}}$ 
of the models, we use the fact 
that ${\mathcal F}$ is unitarily 
equivalent to the constant fiber 
direct integral $L^{2}({\mathbb R}^{d}; 
{\mathcal F}_{\mathrm b})$, 
i.e., 
$${\mathcal F} \equiv L^{2}({\mathbb R}^{d})\otimes 
{\mathcal F}_{\mathrm b} 
\cong  L^{2}({\mathbb R}^{d} ; {\mathcal F}_{\mathrm b}) 
\equiv \int^{\oplus}_{{\mathbb R}^{d}}
{\mathcal F}_{\mathrm b}dx$$ 
(see \cite{rs4,schmuedgen}). 
Throughout this section, we identify ${\mathcal F}$ 
to the constant fiber direct integral, 
i.e., 
\begin{eqnarray*}
{\mathcal F} = \int_{{\mathbb R}^{d}}^{\oplus}
{\mathcal F}_{\mathrm b}dx. 
\label{eq:identification-fiber}
\end{eqnarray*}

If a measurable function $\rho(k)$ 
satisfy $1^{<\Lambda}\rho \in L^{2}({\mathbb R}^{d})$, 
we give the interaction Hamiltonian $H_{\mathrm{I}}$ 
by the so-called Fr\"{o}hlich interaction \cite{hfroehlich}:  
\begin{eqnarray*} 
H_{{\mathrm{I}}} := 
\textsl{q}\int^{\oplus}_{{\mathbb R}^{d}}
\bigl\{
a(1^{<\Lambda}\rho e^{-ikx}) 
+ a^{\dagger}(1^{<\Lambda}\rho e^{-ikx})
\bigr\}
dx   
\label{eq:interaction-Hamiltonian}
\end{eqnarray*}
for every $\textsl{q} \in  \mathbb{R}$. 
Symbolically using the kernels of the annihilation 
and creation operators, the interaction Hamiltonian 
$H_{{\mathrm{I}}}$ is often expressed as 
$$H_{{\mathrm I}}  
= 
\textsl{q}\int_{|k| < \Lambda} 
\left( \rho(k)e^{ikx}a(k) 
 + \overline{\rho(k)}e^{-ikx}
a^{\dagger}(k)\right)dk.   
$$

We also assume the following: 
\begin{description}
\item[(N2)] $1^{<\Lambda}\rho$,\, 
$1^{<\Lambda}\rho/\sqrt{\omega} \in  
L^{2}(\mathbb{R}^{d})$.
\end{description}

The Hamiltonian $H_{\mbox{\rm {\tiny QFT}}}$ 
of the models we consider in this section 
is given by 
\begin{eqnarray*}
H_{\mbox{\rm {\tiny QFT}}} 
&:= & 
H_{0}  + H_{{\mathrm I}} \\ 
&=&  
H_{\mathrm {at}}\otimes I 
+ I\otimes H_{\mathrm{b}} 
+ \textsl{q}\int^{\oplus}_{{\mathbb R}^{d}}
\bigl\{
a(1^{<\Lambda}\rho e^{-ikx}) 
+ a^{\dagger}(1^{<\Lambda}\rho e^{-ikx})
\bigr\}
dx
\end{eqnarray*}
acting in ${\mathcal F}$ \cite{llp,lp}. 
Then, we call this $H_{\mbox{\rm {\tiny QFT}}}$ 
the \textit{Lee-Low-Pines (LLP) Hamiltonian} in this paper, 
though it is called the Pauli-Fierz Hamiltonian 
in \cite{dg,ggm,gerard}. 
Because we are interested in the models in 
solid state physics.

As explained in \cite{hirokawa-nelson}, 
we have the following assertion:

\begin{proposition} 
\label{proposition:nelson-sa}
Assume (N1) and (N2). Then, 
$H_{\mbox{\rm {\tiny QFT}}}$ is self-adjoint with 
$D(H_{\mbox{\rm {\tiny QFT}}})$ $=$ $D(H_{0})$ 
$\equiv$ $D(H_{\mathrm {at}}\otimes I)
\cap D(I\otimes d\Gamma(1))$. 
$H$ is bounded from below for arbitrary 
values of $\textsl{q}$. 
\end{proposition}

Once we assume the existence of a ground state, 
it has to have the property of the spatial 
localization as stated in 
Propositions \ref{proposition:localization} and 
\ref{proposition:exponential-decay} below.  

In the same way as in 
\cite[Proposition 6.1]{hirokawa-nelson} 
we can prove the following: 

\begin{proposition}
\label{proposition:localization}
Assume (N1-1) and (N2). 
If $H_{\mbox{\rm {\tiny QFT}}}$ has a ground state 
$\Psi_{\mbox{\rm {\tiny QFT}}}$, then 
$\Psi_{\mbox{\rm {\tiny QFT}}} \in  
D(x^{2}\otimes I)$. 
\end{proposition}

In the same way as in \cite[Proposition 6.3]{hirokawa-nelson}, 
obeying the idea in \cite{gll} with a little modification 
to meet our models, we have the following:     

\begin{proposition}
\label{proposition:exponential-decay}
Assume (N1-2) and (N2). 
If $H_{\mbox{\rm {\tiny QFT}}}$ has a ground state 
$\Psi_{\mbox{\rm {\tiny QFT}}}$, 
then there is $C_{0} > 0$ such that 
$\Psi_{\mbox{\rm {\tiny QFT}}} \in  D(e^{C_{0}|x|})$. 
\end{proposition}

\begin{remark}
\label{rem:UC}
Propositions \ref{proposition:localization} 
and \ref{proposition:exponential-decay} 
tell us that if LLP Hamiltonian has a ground state, 
the uncertainty $\Delta_{\mathrm{gs}}|x| 
:= \langle (|x|-\langle |x|\rangle_{\mathrm{gs}})^{2} 
\rangle_{\mathrm{gs}}^{1/2}$ 
of $|x|$ in the ground state is finite. 
This is a natural fact in quantum theory 
to observe the electron in the ground state. 
On the other hand, Theorem \ref{corollary:absence2-2} 
tells us that IR singularity condition causes 
IR catastrophe for LLP Hamiltonian. 
Thus, since the electron has to dress itself in 
the cloud of infinitely many soft bosons, 
we can expect that \cite[Theorem 2.1]{hirokawa-nelson} 
also holds for LLP Hamiltonian. 
Namely, $\Delta_{\mathrm{gs}}|x|$ must diverge. 
Therefore, supposing the existence of a ground state 
under IR singularity condition brings about a contradiction 
in quantum theory. 
We find this contradiction in a logic of mathematics 
to show Theorem \ref{theorem:criterion} below.   
\end{remark}

The following OPPT formula can be proved 
in the same way as in 
\cite[Proposition 3.1]{hirokawa-nelson}:  

\begin{proposition} 
\label{proposition:pull-through-formula}
Assume (N1) and (N2). Then, 
for all $f \in$  $C_{0}^{\infty}$ 
$({\mathbb R}^{d}
\setminus \left\{ 0\right\})$, 
\begin{eqnarray*} 
a(f)\Psi_{\mbox{\rm {\tiny QFT}}}   
=  
-\, \textsl{q}
\int_{{\mathbb R}^{d}}
\overline{f(k)}\rho(k)
\left(\widehat{H}_{\mbox{\rm {\tiny QFT}}} 
+ \omega(k)\right)^{-1}
e^{-ikx}
\Psi_{\mbox{\rm {\tiny QFT}}}dk,   
\label{eq:pull-through-formula-rigorous'}
\end{eqnarray*}
provided that $\Psi_{\mbox{\rm {\tiny QFT}}} 
\in  D(x^{2}\otimes I)$. 
Therefore,   
\begin{eqnarray*} 
B_{\mbox{\rm {\tiny PT}}}(k) = 
\textsl{q}
1^{<\Lambda}(k)\rho(k)e^{-ikx}\otimes I.   
\label{eq:OPPT-H'}
\end{eqnarray*}
\end{proposition}

To consider the problem mentioned in 
Section \ref{section:Intro} 
(i.e., the problem 
stated in \cite[Remark 2]{hirokawa-nelson}), 
we give an example of 
$\omega(k)$ and $\rho(k)$ here:  

\begin{example}
\label{example:dame}
As an example of the dispersion 
relation $\omega(k)$ and the interaction 
function $\rho(k)$, through a simplification and 
an idealization, let us set them as 
\begin{eqnarray}
\omega(k) = |k|^{\mu}\,\,\,\mbox{and}\,\,\, 
\rho(k) = |k|^{- \nu}
\label{eq:concrete-case}
\end{eqnarray}
for $\mu \ge 0$ and $\nu \in  \mathbb{R}$, respectively. 
Because we are interested in 
IR situation around $k=0$. 
Then, we have 
\begin{eqnarray}
\gamma_{\mathrm{c}}^{} 
= \frac{d - 2\nu}{2\mu}.
\label{eq:gamma-c}
\end{eqnarray} 
Here we note we can consider $\gamma_{\mathrm{c}}^{}$ 
to be infinite when $\mu=0$ because (N2) requires that 
$\nu$ should be less than $d/2$ (i.e., $\nu < d/2$) 
and thus all $\gamma$ are in the IR-safe region 
in this case. 
The condition, $d \le 2(\mu + \nu)$, 
implies $\lambda/\omega \notin  
L^{2}(\mathbb{R}^{d})$. 
A sufficient condition so that 
we can obtain $\gamma_{0}^{}$ in (2) 
of Corollary \ref{proposition:absence1-1} and  
Eq.(\ref{eq:DG}) 
is $d > 2(\mu + \nu - 1)$ 
as shown in the proof of (iii) of Theorem 
\ref{theorem:criterion}. 
So, because $H_{\mbox{\rm {\tiny QFT}}}$ 
should be defined to be self-adjoint, 
a sufficient condition so that Corollary 
\ref{proposition:absence1-1} 
works is 
\begin{eqnarray}
\max\left\{ 
\frac{\mu}{2} + \nu\, ,\, 
\mu + \nu - 1\right\} 
< \frac{d}{2} \le  \mu + \nu. 
\label{eq:restriction}
\end{eqnarray}
\end{example}

\qquad 

As in Example \ref{example:dame} 
the dimension $d$ has haven 
a restriction from below if we use 
Corollary \ref{proposition:absence1-1}. 
However, since $(\mu/2) + \nu < \mu + \nu -1$ 
iff $\mu > 2$, there is a possibility that 
$(\mu/2) + \nu < d/2 \le \mu + \nu -1$ 
when $\mu > 2$. 
Thus, Corollary \ref{proposition:absence1-1} 
does not work in this case. 
We try to remove this restriction 
in the case $\mu > 2$ by using 
Theorem \ref{corollary:absence2-3} from now on.

Let us take $\mu$ and $\gamma$ with 
$2< \mu$ and $0 < \gamma < 1 - (2/\mu)$ now. 
If $\nu$ satisfies  
\begin{eqnarray}
\frac{d}{2} - 
\frac{1+\gamma}{2}\mu \le 
\nu 
< 
\min\left\{
\frac{d+1}{2} - \frac{1+\gamma}{2}\mu\, ,\, 
\frac{d}{2}
\right\}, 
\label{eq:condition-nu}
\end{eqnarray}
then we have 
$$
\gamma\mu + \nu 
< \frac{1+\gamma}{2}\mu + \nu - \frac{1}{2} 
< \frac{d}{2} 
\le 
\frac{1+\gamma}{2}\mu + \nu 
< \mu + \nu -1.
$$
Namely, $d, \mu$ and $\nu$ 
are out of the region (\ref{eq:restriction}) 
under Eq.(\ref{eq:condition-nu}). 
But we have the following criterion:

\begin{theorem}(Criterion for IR Catastrophe)
\label{theorem:criterion}
Set $\omega(k)$ and $\rho(k)$ 
as Eq.(\ref{eq:concrete-case}). 
Assume that $\mu + 2\nu < d$.  
Let $V$ is in class (N1) and (N2). 
Then, the following \rm{(i) -- (iv)} hold: 
\begin{description}
\item[(i)] If $\nu + \mu < d/2$, 
then IR catastrophe does not occur, and moreover, 
there is a constant $\textsl{q}_{0}^{} 
\in  \mathbb{R}\cup\{\infty\}$ 
such that ground state exists in ${\mathcal F}$ 
for every $\textsl{q}$ with $|\textsl{q}| < 
\textsl{q}_{0}^{}$. 
\item[(ii)] If $\nu + \mu \ge d/2$, 
then IR catastrophe occurs. 
\item[(iii)] If $d, \mu, \nu$ satisfy 
Eq.(\ref{eq:restriction}), 
then there is no ground state in ${\mathcal F}$.
\item[(iv)] Set $\gamma 
:= 2\gamma_{\mathrm{c}}^{} -1$, where 
$\gamma_{\mathrm{c}}^{}$ is in Eq.(\ref{eq:gamma-c}). 
If $\mu > 2$ and 
$(d/2) - \mu < \nu < d/2$, 
then Eq.(\ref{eq:condition-nu}) holds and 
there is no ground state in ${\mathcal F}$.
\end{description}
\end{theorem}

\demo 
We note the condition, $\mu + \nu < d/2$, 
implies $\lambda/\omega \in L^{2}(\mathbb{R}^{d})$. 
Hence $\| K_{\mbox{\rm {\tiny PT}}}(\cdot)
\|_{\mathcal{F}}  \in L^{2}(\mathbb{R}^{d})$ 
follows from this condition. 
Thus, Theorem \ref{theorem:CNB} tells us that 
$\Psi_{\mbox{\rm {\tiny QFT}}} 
\in  D(N^{1/2})$ if $\Psi_{\mbox{\rm {\tiny QFT}}}$ exists. 
Namely, IR catastrophe does not occur. 
The existence of a ground state 
$\Psi_{\mbox{\rm {\tiny QFT}}}$ 
is due to Spohn's result \cite{spohn}. 
Thus, part (i) is completed. 

Part (ii) follows from Theorem \ref{corollary:absence2-2}. 

To prove part (iii) we use the reduction of absurdity. 
Suppose that there is a ground state 
$\Psi_{\mbox{\rm {\tiny QFT}}}$. 
The inequality $d/2 < \mu+\nu$ in Eq.(\ref{eq:restriction}) 
implies that 
$\gamma_{\mathrm{c}}^{} \equiv (d-2\nu)/2\mu <1$, 
so $1$ is in the IR-divergent region. 
Moreover, we have 
$1 -\mu^{-1} < \gamma_{\mathrm{c}}^{}$ 
by $\mu+\nu-1 < d/2$ in Eq.(\ref{eq:restriction}). 
Thus, every $\gamma_{0}^{}$ with $1-\mu^{-1} 
\le \gamma_{0}^{} < \gamma_{\mathrm{c}}^{}$ 
is in the IR-safe region.   
Thus, taking $g(k) = \textsl{q}\lambda(k) = 
\textsl{q}1^{<\Lambda}(k)\rho(k)$, 
the assumptions (1) and (2) of 
Corollary \ref{proposition:absence1-1} hold. 
Taking $J_{\mathrm{err}}(k) = \textsl{q}\lambda(k)
(e^{-ikx} - 1)\otimes I$, 
the assumption (3) of Corollary \ref{proposition:absence1-1} 
holds. 
Moreover, since $1-\mu^{-1}\le\gamma_{0}^{}$ implies 
$(\gamma_{0}^{}-1)\mu + 1 \ge 0$, we have  
\begin{eqnarray*}
&{}& 
\sup_{k\in\mathbb{R}^{d}}\omega(k)^{\gamma_{0}^{}-1}
g(k)^{-1}\| J_{\mathrm{err}}(k)\Psi_{\mbox{\rm {\tiny QFT}}}
\|_{\mathcal{F}} \\ 
&=& \sup_{k\in\mathbb{R}^{d}}\omega(k)^{\gamma_{0}^{}-1}
\| \left((e^{-ikx} - 1)\otimes I\right)\Psi_{\mbox{\rm {\tiny QFT}}}
\|_{\mathcal{F}} \\ 
&\le&  
\left( 
\sup_{k\in\mathbb{R}^{d}} 
|k|^{(\gamma_{0}^{}-1)\mu + 1}
\right)
\| \left(|x|\otimes I\right)\Psi_{\mbox{\rm {\tiny QFT}}}
\|_{\mathcal{F}} \\ 
&\le& 
\| \left(|x|\otimes I\right)\Psi_{\mbox{\rm {\tiny QFT}}}
\|_{\mathcal{F}} 
< \infty
\end{eqnarray*}
by Propositions \ref{proposition:localization}, 
\ref{proposition:exponential-decay} 
and \ref{proposition:pull-through-formula}.  
This contradicts the assertion of 
Corollary \ref{proposition:absence1-1}. 

Using the reduction of absurdity, 
we prove part (iv). 
Thus, we suppose there is a ground 
state $\Psi_{\mbox{\rm {\tiny QFT}}}$.  
Our assumption of (iv) yields 
Eq.(\ref{eq:condition-nu}) immediately. 
It is clear that $\gamma$ is in the IR-safe region 
and $(1+\gamma)/2$ in the IR-divergent region. 
Set $\lambda(k)$ and $B_{0}(k)$ 
as $\lambda(k) = 1^{<\Lambda}(k)\rho(k)$ 
and $B_{0}(k) = \textsl{q}e^{-ikx}\otimes I$ 
respectively. 
Then, $B_{\mbox{\rm {\tiny PT}}}(k) 
= \lambda(k)B_{0}(k)$ by Propositions 
\ref{proposition:localization}, 
\ref{proposition:exponential-decay}, 
\ref{proposition:pull-through-formula}. 
It is easy to check 
$$\int_{\mathbb{R}^{d}} 
\frac{|k_{j}||\lambda(k)|^{2}}{\omega(k)^{1+\gamma}}dk 
= 
\int_{\mathbb{R}^{d}} 
\frac{|k_{j}||\lambda(k)|^{2}}{
\omega(k)^{2\gamma_{\mathrm{c}}^{}}}dk 
\le 
\int_{\mathbb{R}^{d}}\frac{|\lambda(k)|^{2}}{
\omega(k)^{2(\gamma_{\mathrm{c}}^{}-(1/2\mu))}}
dk 
< \infty 
$$
since $\gamma_{\mathrm{c}}^{}-(1/2\mu)$ is in the 
IR-safe region. 
We note $B_{0}(0)\Psi_{\mbox{\rm {\tiny QFT}}} 
= \textsl{q}\Psi_{\mbox{\rm {\tiny QFT}}} 
\ne 0$. 
Applying Maclaurin's theorem to 
$f(t) := e^{-itkx}$\, ($t \in  [0,1]$), 
there is a $\theta$ with $0 < \theta < 1$ 
such that $B_{j}(k) = -ix_{j}e^{-i\theta kx}\otimes I$. 
Thus, Propositions \ref{proposition:localization} 
and 
\ref{proposition:exponential-decay} 
lead us to the conclusion that 
$\sup_{k\in\mathbb{R}^{d}}
\| B_{j}(k)\Psi_{\mbox{\rm {\tiny QFT}}}
\|_{\mathcal{F}} 
\le 
\| |x|\Psi_{\mbox{\rm {\tiny QFT}}}
\|_{\mathcal{F}} < \infty$. 
However, the last two facts 
contradict the statement of 
Theorem \ref{corollary:absence2-3}.  
\qed

\hfill\break 
{\large {\bf Acknowledgement}} 
 
The author is very grateful to 
L. Bruneau, J. Derezi\'{n}ski, 
C. G\'{e}rard, and F. Hiroshima for useful comments 
and discussions.

\begin{flushleft}
Masao Hirokawa \\ 
Graduate School of Natural Science and Technology, \\ 
Okayama University, \\   
700-8530 Okayama, \\ 
Japan \\ 
e-mail: \verb*|hirokawa@math.okayama-u.ac.jp|
\end{flushleft}

\end{document}